\documentclass[a4paper,10pt]{article}

\def\fnote#1#2{\begingroup\def\thefootnote{#1}\footnote{#2}\addtocounter
{footnote}{-1}\endgroup}

\usepackage{amsfonts}
\usepackage{amssymb}
\usepackage{graphics}
\usepackage{epsfig}
\usepackage{graphicx}
\usepackage{epsfig}
\usepackage{amssymb}
\usepackage{amsfonts}

\newtheorem{theorem}{Theorem}

\newtheorem{proposition}[theorem]{Proposition}

\newenvironment{proof}[1][Proof]{\noindent\textbf{#1.} }{\ \rule{0.5em}{0.5em}}
\newskip\humongous \humongous=0pt plus 1000pt minus 1000pt

\newcommand {\slsh} [1] {\not{\hbox{\kern-2pt${#1}$}}}

\newcommand{\gsim}{\lower.7ex\hbox{$\;\stackrel{\textstyle>}{\sim}\;$}}
\newcommand{\lsim}{\lower.7ex\hbox{$\;\stackrel{\textstyle<}{\sim}\;$}}
\newcommand {\beq} {\begin{equation}}
\newcommand {\eeq} {\end{equation}}
\newcommand {\beqn}{\begin{eqnarray}}
\newcommand {\eeqn} {\end{eqnarray}}
\newcommand{\bea}{\begin{eqnarray}}
\newcommand{\eea}{\end{eqnarray}}

\def\SU{{\rm SU}}
\def\U{{\rm U}}
\def\SO{{\rm SO}}
\def\N{{\cal N}}
\def\Z{\mathbb{Z}}
\def\O{{\cal O}}
\def\AdS{{\rm AdS}}
\def\L{{\rm L}}
\def\R{{\rm R}}

\begin{document}

\hfill{ 
FTPI-MINN-08/46; UMN-TH-2730/08 }

\vspace{36pt}

\begin{center}
{\large { { Skyrmions in Orientifold and Adjoint QCD }}}

\vspace{36pt}
Stefano Bolognesi\fnote{*}{ 
bolognesi@physics.umn.edu}\\
\vspace{20pt}
{\em  William I. Fine Theoretical Physics Institute, University of Minnesota, \\ 116 Church St. S.E., Minneapolis, MN 55455, USA.}
\end{center}

\vspace{16pt}

\begin{abstract}

This contribution is a review of recent developments regarding the Skyrmion sector of higher representation QCD.   This review is mostly based on the results of  \cite{Bolognesi:2006ws,Bolognesi:2007ut,Auzzi:2008hu}.

Ordinary QCD is a $\SU(n)$ gauge theory with $n_f$ Dirac quarks in the fundamental representation. Changing the representation of quarks leads to different and interesting theories, which are not as well studied as ordinary QCD. In order to be able to have a consistent asymptotically free large $n$ limit, we must limit ourselves to three cases: two-index representation (symmetric or anti-symmetric) and adjoint representation. We call the first two  ``orientifold  QCD (S/A)'' and the last one ``adjoint QCD''.

Skyrmions of the low-energy effective Lagrangian shall be the main subject of this review. There are puzzling aspects, both in orientifold and adjoint QCD, regarding the identification of the Skyrmion and its quantum stability, that have not yet been understood.  We shall explain these problems and the solution we proposed for them. 

The first part is dedicated to the two-index (S/A) representation. Here the challenge is to identify the correct particle in the spectrum that has to be identified with the Skyrmion. It turns out {\it not} to be the simplest baryon (as in ordinary QCD) but a baryonic state with higher charge, precisely composed by $n(n\pm 1)/2$ quarks. Although not the simplest among the baryons, it is the one that minimizes the mass per unit of baryonic charge and thus is the most stable among them.

The second part is devoted to the quarks in the adjoint representation.  The task here assume  a different perspective. We still have a Skyrmion, but we do not have a baryon charge, like in ordinary QCD. An important role is now played by a massive fermion that must be considered in the low-energy effective Lagrangian. Through this fermion, the Skyrmion acquires an anomalous fermionic number $(-1)^F$ and, as a consequence, an odd relationship between the latter and its spin/statistic. This implies a $\Z_2$ stability of the Skyrmion.

\end{abstract}

\newpage
\section{Introduction}

This is a story about large $n$, Skyrmions, and baryons.

The large $n$ expansion is a major tool in the study of strongly coupled
$\SU(n)$\ gauge theories \cite{'tHooft:1973jz}. In double line notation,
gluons are represented by two lines with opposite oriented arrows while quarks,
if in the fundamental representation, are represented by a
single oriented line. Every Feynman diagram corresponds to a certain
topological oriented surface with a certain number of handles and holes. Holes
correspond to quark loops. Every handle suppresses the diagram by a factor of
$n^{-2}$ and every hole by a factor of $n^{-1}$. The large $n$ limit is thus
dominated by diagrams with only planar gluons, and fermion quantum effects are
present only in the subleading orders.

In ordinary QCD (quarks in the fundamental representation), of
particular interest for what follows are the baryons, whose gauge wave function
is%
\begin{equation}
\epsilon_{\alpha_{1}\dots\alpha_{n}}\,Q^{\alpha_{1}}\dots Q^{\alpha_{n}} \ .
\label{wave}
\end{equation}
It is a gauge singlet completely antisymmetric under the exchange of any two quarks.
The antisymmetric property of the gauge wave function (\ref{wave}), and the fermionic nature of quarks, implies
that the spatial wave function is symmetric under the exchange of quarks. In the
large $n$ limit, the baryon can be approximated as a system of free bosons
confined in a mean potential. The mass of this Bose-Einstein condensate scales
like the number of particles $n$. In the large $n$ limit, it can be identified with the solitons of the chiral
effective Lagrangian \cite{Skyrme:1961vr,Witten:1983tx}.

Recently, another kind of theories has received considerable attention.
This is the case of quarks in the two-index, symmetric or antisymmetric (S/A),
representation. Armoni, Shifman and Veneziano have shown that a theory
with $n_{f}$ Dirac quarks in the two-index S/A representation is equivalent,
in a certain bosonic subsector and in the large $n$ limit, to a theory with
$n_{f}$ Weyl quarks in the adjoint representation \cite{ASV}.
Particularly interesting is the antisymmetric representation since it can be
used to reproduce QCD at $n=3$.  This equivalence\ becomes particularly
useful when $n_{f}=1$ since the theory with one fermion in the adjoint is
${\cal N}=1$ super Yang-Mills and some non-perturbative results are known
about it.

The large $n$ limit with a fixed number of quarks in the fundamental
representation has the disadvantages that all the quantum corrections due to
quark loops vanish as $1/n$. For example, the $\eta^{\prime}$ mass
vanishes like $1/n$ since its value comes only from the axial
$\U(1)_{\rm A}$ anomaly. That is not true for the large $n$ of orientifold theories.

If the number of flavors $n_f$ is greater than one, the theory under consideration has a
chiral symmetry that is spontaneously broken by the quark condensate.  The pattern of the chiral
symmetry breaking ($\chi$SB)   is identical to that of QCD, namely,
\begin{equation}
\SU(n_f)_{\rm L} \times \SU(n_f)_{\rm R} \to \SU(n_f)_{\rm V} \ .
\end{equation}
As a result, the low-energy limit is described by the same chiral Lagrangian
as in QCD, with the only distinction that in the case of two-index
quarks the pion constant $F_\pi$ scales as
\beq
F_\pi^2 \sim n^2\,,
\eeq
while in QCD it scales like  $F_\pi^2 \sim n$.

As was noted in
\cite{Armoni:2003jk}, this seemingly minor difference leads to a crucial
consequence. In both theories there exist Skyrmion solitons. The very same
dependence of the Skyrmion mass on $F_\pi^2$ in the theory with the
two-index antisymmetric quarks implies that the Skyrmion mass grows as $n^2$
rather than $n$.
At first sight, this is totally counterintuitive since in this case the simplest baryon we could imagine is in fact
\begin{equation}
\epsilon_{\alpha_{1}\alpha_{2}\dots\alpha_{n}}\epsilon_{\beta_{1}\beta
_{2}\dots\beta_{n}}\,Q^{\alpha_{1}\beta_{1}}Q^{\alpha_{2}\beta_{2}}\dots
Q^{\alpha_{n}\beta_{n}} \ ,
\end{equation}
and just as
in QCD, it is a color-singlet from $n$ quarks.

This puzzle, and its solution \cite{Bolognesi:2006ws}, will be the content of the first part of the paper.
It turns out that
if one considers $n$-quark colorless bound states, not all quarks can be in
the $S$-wave state, and as a result, the mass of such objects scales faster than expected. The minimal number of quarks of which one
can build a particle with all quarks in the $S$-wave state is $n(n \pm 1)/2$.
This $\sim n^2$ quark particle is the stable state described by the
Skyrmion. As for the $n$-quark particle they are unstable with
respect to fusion of $n$ species into one $\sim n^2$ quark state, with a release of energy in the form of pion emission.

\vskip 0.50cm
\begin{center}
*  *  *
\end{center} 
\noindent
The second part of the paper is devoted to the Skyrmion in adjoint QCD. We
will consider $n_f $ massless \emph{Weyl} (or, which is the same, \emph{
Majorana}) fermions in the adjoint representation of the SU$(n)$ gauge
theory.  The pattern of the $\chi$SB in this case is
\begin{equation}
\mathrm{\SU}(n_f)\times \mathbb{Z}_{2n n_f}\to \mathrm{SO}(n_f)
\times \mathbb{
Z} _{2} \,,  
\end{equation}
where the discrete factors are the remnants of the anomalous singlet axial
U(1).

The low-energy pion Lagrangian is a nonlinear sigma model with the
target space $\mathcal{M}$ given by the coset space
\begin{equation}
\mathcal{M}_{n_f} = \mathrm{SU}(n_{f})/\mathrm{SO}(n_{f}) \ .  \label{targ}
\end{equation}
For the particular case $n_f=2$, it is $\mathcal{M}_2 = \mathrm{SU}(2)/ \mathrm{U}(1) =S^2$.
The soliton's topological
stability is due to the existence of the Hopf invariant or, equivalently, the fact that 
the third
homotopy group for (\ref{targ}) is nontrivial.

Unlike ordinary and orientifold QCD, where the relationship between
the Skyrmions and microscopic theory is related to the conserved baryon number, in this case the relationship is
far from clear. 
Is  the Skyrmion stability
an artifact of the low-energy approximation? If no, what prevents these
particles whose mass scales as $n^2$ from decaying into ``light"
color-singlet mesons and baryons with mass $\O(n^0)$?  These questions will be our main concern.

We shall find that the Skyrmion is indeed stable, due to an anomalous relation between its spin/statistic and the fermion number. 
A key role in this analysis is played by a composite fermion, with mass $\O(n^0)$ that must be added to the low-energy effective Lagrangian to have a complete account of the spectrum of the theory.

\section{Orientifold  QCD}
\label{twoindex}

We now consider orientifold QCD.  It consists of $\SU(n)$ gauge theory with $n_{f}$ Dirac
fermions transforming according to the two-index symmetric or antisymmetric representation,
\begin{equation}
\mathcal{L}=-\frac{1}{2}\mathrm{Tr~}F_{\mu\nu}F^{\mu\nu}+\sum_{f=1}^{n_{f}%
}\overline{Q}^{f}(iD_{\mu}\gamma^{\mu}-m_{f})Q_{f} \ ,\label{lagrangian}
\end{equation}
where $f$ is a flavor index and the gauge indices are suppressed.
The field strength is $F_{\mu\nu}=\partial_{\mu}A_{\nu}-\partial_{\nu}A_{\mu}+ig[A_{\mu},A_{\nu}]$, $g$ is the coupling constant, and the covariant
derivative is $D_{\mu}=\partial_{\mu}-igA_{\mu}$. 
We use the conventions to indicate $Q^{\{\alpha\beta\}}$
 a quark in the two-index symmetric representation while
$Q^{[\alpha\beta]}$  a quark in the two index antisymmetric
representation. When we write $Q^{\alpha\beta}$, it means that what we are writing is valid for both representations.

This part is organized as follows. In Section \ref{largen}, we briefly recall some basic feature of the large $n$ limit. In Section \ref{effective}, we study effective Lagrangian and the Skyrmion properties. In Section
\ref{puzzle}, we study baryons at large $n$. In Section
\ref{stablebaryons} we find the stable baryons that can be identified with the
Skyrmions in the large $n$ limit. In Section \ref{moreantisymmetric}, we
consider some peculiar property of the antisymmetric representation. Finally
in Section \ref{stabilitytwo} we discuss the stability of the Skyrmion.

\subsection{Large $n$ Limit}
\label{largen}

\begin{figure}
[h!t]
\begin{center}
\leavevmode \epsfxsize 6.5 cm \epsffile{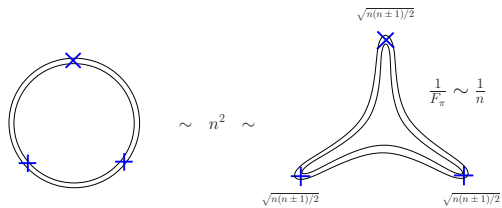}
\caption{{\protect\footnotesize The $n$  dependence of the meson
coupling $F_{\pi}$.}}%
\label{effepai}%
\end{center}
\end{figure}

\begin{figure}
[h!t]
\begin{center}
\leavevmode \epsfxsize 6.5 cm \epsffile{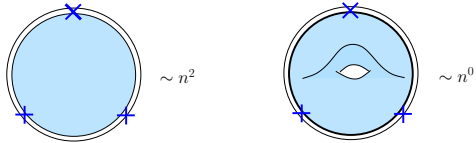}
\caption{\footnotesize First-order and second-order
contributions to
the three-meson interaction.}
\label{torus}
\end{center}
\end{figure}

In order to have a well-defined large $n$ limit, we take the product $g^{2}\,n$ to be finite.
At large $n$, the theory reduces to an infinite tower of weakly coupled hadrons
whose interaction strength vanishes like $n^{-2}$. The large $n$ behavior of
orientifold QCD is very similar to that of theories with fermions in the adjoint
representation. The dependence upon the number of colors of
the meson coupling can be evaluated using the planar diagrams presented in
Figure \ref{effepai} and paying attention to the hadron wave function
normalization.
We will denote the decay constant of the typical meson by $F_{\pi}$.
Using a double line notation, the Feynman diagrams can be arranged according to
the topology of the surface related to the diagram. The $n$ powers of the
Feynman diagrams can be read off from two topological properties of the
surface: the number of handles and the number of holes. Every handle carries a
factor $n^{-2}$, and every hole carries a factor $n^{-1}$. In the ordinary 't
Hooft limit, where the quarks are taken in the fundamental representation, the
holes are given by the quark loops. In the higher representation case, quarks
are represented by double lines as the gluons and so there are no holes. The contribution to $F_{\pi}$ in the large $n$ limit can thus be
arranged as in Figure \ref{torus} where the leading order is a quark closed
double line with planar quarks and gluons inside, and the next subleading
order is given by adding a handle. The leading order scales like $n^{2}$ while
the subleading order scales like $n^{0}$.
The previous color counting is not affected by the addition of a finite number
of flavors.

\subsection{Effective Lagrangians, Anomalies, and Skyrmions}
\label{effective}

With $n_{f}$ {\it massless} flavors, the Lagrangian (\ref{lagrangian}) has
global symmetry $\SU(n_{f})_{\L}\times \SU(n_{f})_{\R}$. The global chiral
symmetry is expected to dynamically break to its maximal diagonal subgroup by the quark condensate.
The low-energy effective
Lagrangian describes the dynamics of the massless mesons that are the
Nambu-Goldstone bosons of the spontaneous chiral symmetry breaking. Written in
terms of the matrix $U(x)=\exp\left(i \pi(x) / F_{\pi}\right)  $,
where $\pi(x)$ is the Goldstone boson matrix, the effective Lagrangian is%
\begin{equation}
S_{\mathrm{eff}}=\frac{1}{16}F_{\pi}^{2}\int d^{4}x\,\left\{  \mathrm{Tr}%
\partial_{\mu}U\partial_{\mu}U^{-1}+\mathrm{higher~derivatives}\right\}
+ k \Gamma_{\mathrm{WZNW}} \ .
\end{equation}
The topological Wess-Zumino-Novikov-Witten (WZNW) term $\Gamma_{\mathrm{WZNW}}$ is crucial in order
to satisfy the 't Hooft anomaly conditions at the effective Lagrangian level.
Gauging the WZNW term with  respect to the electromagnetic interactions yields
the familiar $\pi^{0} \rightarrow 2\gamma$ anomalous decay. The WZNW term can be
written as
\beq
\Gamma_{\mathrm{WZNW}}=-\frac{i}{240\pi^{2}}\,\int_{\mathcal{M}^{5}}%
\epsilon^{\mu\nu\rho\sigma\tau}\mathrm{Tr}\left(  \partial_{\mu}%
UU^{-1}\partial_{\nu}UU^{-1}\partial_{\rho}UU^{-1}\partial_{\sigma}%
UU^{-1}\partial_{\tau}UU^{-1}\right)  \ .
\eeq
where the integral must be performed over a five-dimensional manifold whose
boundary is ordinary Minkowski space. Quantum consistency of the theory
requires $k$ to be an integer.  Matching with the underlying anomaly
computations requires $k$ to be equal to the number of quarks with respect to
the color. In the case of the fundamental representation $k=n$ while for orientifold QCD, $k=n(n\pm1)/2$.

The low-energy effective theory supports solitonic excitations. In order to obtain
classically stable configurations, it is necessary to include at least a four-derivative term in addition to the usual two-derivative term. 
Higher-derivatives terms are certainly present in the low-energy effective Lagrangian and are crucial for  the Skyrmion stability.

The Skyrmion is a texture-like solution of the effective Lagrangian
arising from the nontrivial third homotopy group of the possible
configurations of the matrix $U(x)$ (namely $\pi_{3}\left(  \SU(n_{f})\right)
=\Z$). In the large $n$ limit, we can treat the effective Lagrangian as
classical, and, since the $n$ dependence appears only as a multiplicative
factor, the size and the mass of the Skyrmion scale, respectively, as $n^{0}$
and $n(n\pm1)/2$. Following \cite{Witten:1983tx}, we can read off the statistics and the
baryon number of the Skyrmion from the coefficient of the WZNW term. The baryon
number of the Skyrmion is $n(n\pm1)/2$ the baryon number of the quarks,
and the statistic is fermionic or bosonic accordingly if $n(n\pm1)/2$
is odd or even.

The results we have just obtained all point in the same direction. There
should exist in the spectrum of the theory a stable baryon that in the large
$n$ limit could be identified with the Skyrmion. This baryon should be
constituted by $n(n\pm1)/2$ quarks, and its mass should scale as
$n^{2}$ in the large $n$ limit.

\subsection{The Puzzle}
\label{puzzle}

Now we introduce the problem we are going to face.

It has been noted in
\cite{Armoni:2003jk} that, at least at  first glance, the identification
between baryons and Skyrmions in the large $n$ limit, for orientifold QCD, is problematic. 
A natural
choice for the wave function of the baryon is the following
\begin{equation}
\label{naivebaryon}
\epsilon_{\alpha_{1}\alpha_{2}\dots\alpha_{n}}\epsilon_{\beta_{1}\beta
_{2}\dots\beta_{n}}\,Q^{\alpha_{1}\beta_{1}}Q^{\alpha_{2}\beta_{2}}\dots
Q^{\alpha_{n}\beta_{n}} \ . \label{firstguess}
\end{equation}
This baryon is formed of $n$ quarks and two epsilon tensors to saturate the indices. The first guess, since the number of components is $n$, is
that its mass scales like $n$ in the large $n$ limit. The mass of the
Skyrmion scales instead similar to $F_{\pi}^{2}$ that, in the case of the quarks in higher representations, is $n^{2}$.  This is the first discrepancy between the baryon (\ref{firstguess}) and the Skyrmion.

Let us remember, for a moment, the well-known case of ordinary QCD.
We briefly consider the large $n$ behavior of the baryon in ordinary QCD.
The gauge wave function is%
\begin{equation}
\epsilon_{\alpha_{1}\dots\alpha_{n}}\,Q^{\alpha_{1}}\dots Q^{\alpha_{n}}~,
\label{ordinarybaryon}%
\end{equation}
and it is antisymmetric under the exchange of any two quarks.  Since the quarks are
fermions, the total gauge function $\psi_{\mathrm{gauge}}\psi
_{\mathrm{spin/flavor}}\psi_{\mathrm{space}}$ must be antisymmetric under the
exchange of two quarks. The simplest choice is to take a completely symmetric
spin wave function and a completely symmetric spatial wave function.%
\begin{equation}%
\begin{tabular}
[c]{ccc}%
$\psi_{\mathrm{gauge}}$ & $\psi_{\mathrm{spin/flavor}}$ & $\psi
_{\mathrm{space}}$\\
$-$ & $+$ & $+$%
\end{tabular}
\end{equation}

In the large $n$ limit, the problem can be approximated by a system of free
bosons in a mean field potential $V_{\mathrm{mean}}\left(  r\right)  $ created
by the quarks themselves. The ground state is a Bose-Einstein condensate; the
quarks are all in the ground state of the mean field potential. The large $n$
behavior of the baryon is the following%
\begin{equation}
R\sim\mathcal{O}\left(  1\right)  ~,\qquad M\sim\mathcal{O}\left(  n\right) \ ,
\end{equation}
where $R$ is the size of the baryon and $M$ its mass.

The key point to obtain this result is that the many body problem becomes
enormously simplified by the fact that the coupling constant scales like
$1/g^{2} \sim n$\ in the large $n$ limit. To find the mass in this many
body problem, we have to sum up all the contributions from $k$-body
interactions. The one-body contribution is simply $n$ times the mass of the
single quark. The two-body interaction is of order $1/n$ but
an additional combinatorial factor ${n \choose 2}$ is needed, and we obtain a
contribution to the energy of order $n$. In general, any $k$-body interaction
is of order $1/n^{k-1}$ in the planar limit, and multiplied by
the combinatorial factor ${n \choose k}$, it gives a contribution of order $n$.
The same argument implies that the mean field potential $V_{\mathrm{mean}}(r)$
is constant in the large $n$ limit and so also the typical size of baryon $R$ (roughly the width of the ground state wave function).

These arguments are consistent with the low-energy effective Lagrangian point
of view. This Lagrangian is $L_{\rm eff}\sim n\left(  \partial U\partial
U+\partial U\partial U\partial U\partial U+\dots\right)  $ where$\ U$ is a
$\SU(n_{f})$ matrix . Since $n$ is an overall multiplicative factor, the radius
of the Skyrmion is of order one while its mass is of order $n$.

Now let us go back to orientifold QCD.
The first step toward the solution of the puzzle is to realize that the naive expectation
that the mass of (\ref{firstguess}) scales like $n$ is not correct. The reason
is the following. The gauge wave function (\ref{firstguess}) is symmetric
under the exchange of two quarks. Since the total wave function must be
antisymmetric, this means that the space wave function must be antisymmetric. 
The large $n$ baryon must
thus be approximated as a set  of free {\it fermions} in a mean field potential.
Since fermions cannot all be in the same ground state, there is an extra term
in the energy coming from the Fermi zero temperature pressure. At this point,
one could hope that this extra term could compensate for the mismatch and make the
baryon mass scale like $n^{2}$. A more detailed analysis shows that this is not true.

In higher-representations QCD, the simplest baryon is (\ref{naivebaryon}),
If we exchange two quarks, say for example, $Q^{\alpha_{1}\beta_{1}}$ and
$Q^{\alpha_{2}\beta_{2}}$, this is equivalent to the exchange of
$\alpha_{1}\alpha_{2}$ in $\epsilon_{\alpha_{1}\alpha_{2}\dots\alpha_{n}}$ and
$\beta_{1}\beta_{2}$ in $\epsilon_{\beta_{1}\beta_{2}\dots\beta_{n}}$. The
result is that the gauge wave function is symmetric under the exchange of two
quarks. This means that in order to have a total wave function that is
antisymmetric under  the exchange, the spatial wave function $\psi_{\mathrm{space}%
}$ must be antisymmetric.%
\begin{equation}%
\begin{tabular}
[c]{ccc}%
$\psi_{\mathrm{gauge}}$ & $\psi_{\mathrm{spin/flavor}}$ & $\psi
_{\mathrm{space}}$\\
$+$ & $+$ & $-$%
\end{tabular}
\end{equation}

In the large $n$ limit, the problem can be approximated by a system
of free fermions in a mean field potential $V_{\mathrm{mean}}(r)$.
The ground state is a degenerate Fermi gas and is obtained by
filling all the lowest energy states of the mean field potential up
the Fermi surface. Now there are two kind of forces that enter in
the game:
\begin{itemize}
\item[1)] Gauge forces scales like $n$ and are both repulsive and attractive,
\item[2)] Fermi zero temperature pressure scales like $n^{4/3}$ and is only
repulsive.
\end{itemize}
We can thus immediately infer the following that the simplest baryon cannot be
matched with the Skyrmion; this is because the mass of the Skyrmion goes like
$n^{2}$ while the mass of this baryon obviously cannot go faster than
$n^{4/3}$.

Another discrepancy for the candidate baryon (\ref{firstguess}) comes from the
WZNW term of the effective Lagrangian.  From this term, we can read off the
statistics and the baryon number of the Skyrmion. The baryon number is
$n(n\pm1)/2$, where $\pm$ stand, respectively, for symmetric and
antisymmetric representation, and the statistic is fermionic or bosonic
accordingly if $n(n\pm1)/2$ is odd or even. There is no way to recover
this number from the baryon (\ref{firstguess}).

The topological stability of the Skyrmion in the effective Lagrangian
indicates that, at least in the large $n$ limit, there should exist a stable
state composed by $n(n\pm1)/2$ quarks and whose mass scales like
$n^{2}$. This is possible if there exists a color singlet wave function that
not only is composed by $n(n\pm1)/2$ quarks  but is also completely
antisymmetric under the exchange of them. In what follows we are going to show that this
function exists and that $n(n\pm1)/2$ is the {\it unique} number of quarks
needed for its existence. This shall also confirm the stability of these Skyrmions-baryons. In
fact, any baryonic particle with a smaller number of quarks must have the extra
contributions to its mass coming from the spatial Fermi statistics.

\subsection{The Matching of the Skyrmion}
\label{stablebaryons}

We have seen in the previous section that the simplest baryon (\ref{firstguess}) has a gauge wave
function that is symmetric under the exchange of two quarks. This has a drastic
consequence on its mass-vs.-$n$ behavior. In the
following, we will construct the only possible gauge wave function that is
completely antisymmetric under the exchange of any two quarks. We will find that the
required number of quarks, as expected from the Skyrmion analysis, must be
$n(n\pm1)/2$.

\begin{figure}[h!t]
\begin{center}
\leavevmode \epsfxsize 9.5 cm \epsffile{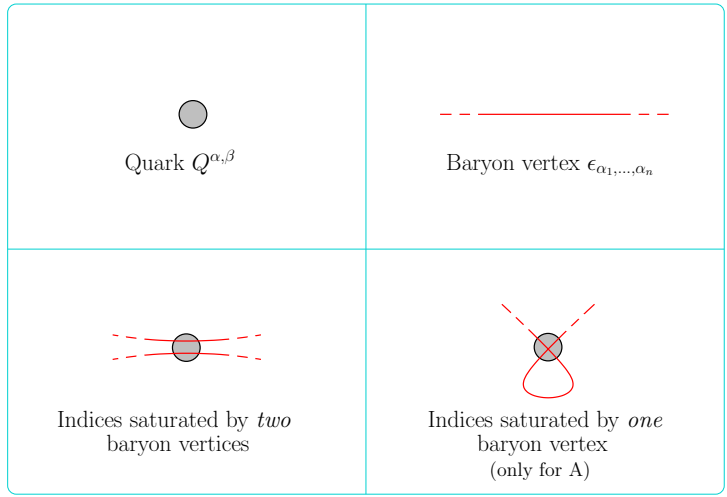}
\end{center}
\caption{\footnotesize  We diagrammatically represent quarks with points and epsilon tensors with lines.
Two lines pass  from every point. Every line connects $n$ points. In the case of the anti-symmetric representation, a line can pass twice on the same point.}
\label{legenda}
\end{figure}

\begin{figure}[h!t]
\begin{center}
\leavevmode \epsfxsize 6.5 cm \epsffile{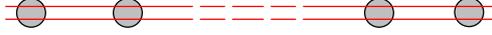}
\end{center}
\caption{\footnotesize Diagrammatic representation for the simplest baryon of Eq.~(\ref{naivebaryon}).}
\label{basic}
\end{figure}

First of all, we introduce a diagrammatic representation of baryons that shall be very useful in the following. As described in Figure \ref{legenda}, we use {\it points} to indicate quarks $Q^{\alpha \beta}$ and {\it lines} to indicate epsilon tensors (or baryon vertices) $\epsilon_{\alpha_1 \dots \alpha_n}$. To have a gauge singlet baryon, we need to build a diagram of points and lines so that:  $1)$ Two lines pass from every point. $2)$ A line passes through $n$ points. Only in the antisymmetric case is it possible for the same line to pass twice on the same point. 
For example, the diagram corresponding to the simplest baryon (\ref{naivebaryon}) is given in Figure \ref{basic}.
We shall now proceed to discuss separately the case of symmetric and antisymmetric representations.

\subsubsection{The symmetric representation}

\begin{figure}[th]
\begin{center}
\leavevmode \epsfxsize 2.2 cm \epsffile{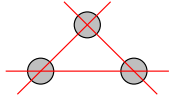}
\end{center}
\caption{\footnotesize Diagrammatic representation for the simplest baryon for $n=2$ (Eq.~(\ref{duesymmetricbaryon})).}
\label{duesymmetric}
\end{figure}

We start from the simplest case: two colors $n=2$. We want to construct a gauge invariant
wave function that contains three quarks $\,Q^{\{\alpha_{1}\beta_{1}\}}$,
$Q^{\{\alpha_{2}\beta_{2}\}}$, and $Q^{\{\alpha_{3}\beta_{3}\}}$, and that is completely
antisymmetric under the exchange of any two of them. A natural guess to try is the triangular diagram of Figure \ref{duesymmetric}.%
The gauge wave function
that corresponds to the diagram is\footnote{Note in particular that the symmetric
representation for $\SU(2)$ is equivalent to the adjoint representation and a
gauge invariant antisymmetric wave function can easily be written as
$\epsilon_{abc}Q^{a}Q^{b}Q^{c}$ where $a,b,c$ are triplet indices. This wave
function is exactly the same as that of Eq. (\ref{duesymmetricbaryon}).}
\begin{equation}
\epsilon_{\alpha_{2}\alpha_{1}}\epsilon_{\beta_{2}\alpha_{3}}\epsilon
_{\beta_{1}\beta_{3}}\,Q^{\{\alpha_{1}\beta_{1}\}}Q^{\{\alpha_{2}\beta_{2}%
\}}Q^{\{\alpha_{3}\beta_{3}\}}~. \label{duesymmetricbaryon}%
\end{equation}
Let us now prove that this wave function is indeed antisymmetric under the exchange of two quarks.
We can proceed with the following algebraic steps:
\bea
&  ~~~~~~\,\epsilon_{\alpha_{1}\alpha_{2}}\epsilon_{\beta_{1}\alpha_{3}%
}\epsilon_{\beta_{2}\beta_{3}}\,Q^{\{\alpha_{1}\beta_{1}\}}Q^{\{\alpha
_{2}\beta_{2}\}}Q^{\{\alpha_{3}\beta_{3}\}} \nonumber \\
&  =-\epsilon_{\alpha_{2}\alpha_{1}}\epsilon_{\beta_{1}\alpha_{3}}%
\epsilon_{\beta_{2}\beta_{3}}\,Q^{\{\alpha_{1}\beta_{1}\}}Q^{\{\alpha_{2}%
\beta_{2}\}}Q^{\{\alpha_{3}\beta_{3}\}}\nonumber\\
&  =-\epsilon_{\alpha_{2}\alpha_{1}}\epsilon_{\beta_{1}\beta_{3}}%
\epsilon_{\beta_{2}\alpha_{3}}\,Q^{\{\alpha_{1}\beta_{1}\}}Q^{\{\alpha
_{2}\beta_{2}\}}Q^{\{\beta_{3}\alpha_{3}\}}\nonumber\\
&  =-\epsilon_{\alpha_{2}\alpha_{1}}\epsilon_{\beta_{1}\beta_{3}}%
\epsilon_{\beta_{2}\alpha_{3}}\,Q^{\{\alpha_{1}\beta_{1}\}}Q^{\{\alpha
_{2}\beta_{2}\}}Q^{\{\alpha_{3}\beta_{3}\}} \ .
\label{pedestrianproof}
\eea
The first line corresponds to Eq.(\ref{duesymmetricbaryon}) with  quarks $1$ and $2$ exchanged (the exchange is done in the epsilon tensors). 
The three algebraic steps are the following:
\begin{itemize}
\item[(A$\rightarrow$B)] Exchange of $\alpha_{1}$ and $\alpha_{2}$ in the
$\epsilon$  brings a minus factor;
\item[(B$\rightarrow$C)] Renomination of $\alpha_{3}$ with $\beta_{3}$, which
has no consequences;
\item[(C$\rightarrow$D)] Exchange of $\alpha_{3}$ and $\beta_{3}$ in the quark
also has no consequences.
\end{itemize}
In the final line, we recover exactly the wave function (\ref{duesymmetricbaryon}) but with a minus sign in front of it.

We now prove this general theorem that, apart from the confirmation of the existence and uniqueness of these $n(n+1)/2$ fully antisymmetric gauge wave function, will also give the recipe to construct it.
\begin{proposition}
\label{proposition1}
\ There is one and only one gauge wave function that is a gauge singlet and
completely antisymmetric under the exchange of two quarks. This wave function is
composed by $n(n+1)/2$ quarks $Q^{\{\alpha\beta\}}$ and is the
completely antisymmetric subspace of the tensor product of $n(n+1)/2$
quarks $Q^{\{\alpha\beta\}}$.
\end{proposition}
\begin{proof}
Call $S$ the number of quarks in a hypothetical gauge wave function that
satisfies the previous conditions.  We need two facts to prove the
proposition: ${\it 1)}$ Two indices   $\alpha_{i}$ and $\beta_{i}$ of the same quark $Q^{\{\alpha_{i}\beta_{i}\}}$ {\it cannot}  belong to the same
saturation line since they are symmetric under the exchange; ${\it 2)}$
Two quarks $Q^{\{\alpha
_{i}\beta_{i}\}}$ and $Q^{\{\alpha_{j}\beta_{j}\}}$ can be connected by {\it at most}  one line. The reason is simply that exchanging them  would give a plus sign
instead of the required minus sign. At this point, we are ready to build the fully antisymmetric wave function. We follow Figure \ref{generalsymmetric}.
\begin{figure}[th]
\begin{center}
\leavevmode \epsfxsize 6.5 cm \epsffile{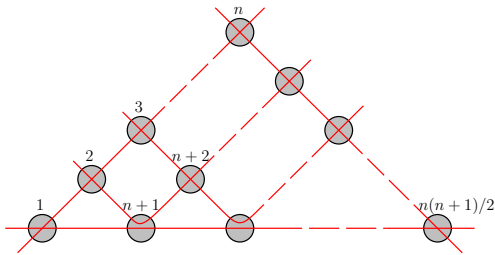}
\end{center}
\caption{\footnotesize The triangular diagram for the Skyrmion-baryon in the case of the two-index symmetric representation: $n(n+1)/2$ points connected by $n+1$ lines.}
\label{generalsymmetric}
\end{figure}
We start from the first quark $Q^{\{\alpha_{1}\beta_{1}\}}$ and draw the first saturation line that departs from this quark. This implies the presence of other $n-1$ quarks that we call $Q^{\{\alpha_{2}\beta_{2}\}} , \dots, Q^{\{\alpha_{n}\beta_{n}\}}$. 
Let's consider now $Q^{\{\alpha_{2}\beta_{2}\}}$. One index is already saturated, and from the other one a new saturation line must depart.  Due to  property ${\it 2)}$, new $n-2$ quarks must be added. One of them,  $Q^{\{\alpha_{n+1}\beta_{n+1}\}}$,  can be in common with $Q^{\{\alpha_{1}\beta_{1}\}}$. We then take $Q^{\{\alpha_{3}\beta_{3}\}}$ and start the other saturation line. It can pass from $Q^{\{\alpha_{n+2}\beta_{n+2}\}}$, but then other different $n-2$ quarks must be added. So on we go until we reach $Q^{\{\alpha_{n}\beta_{n}\}}$ and complete the saturation adding the last line. In total we have $n + n-1 + n-2 +\dots +1 = n(n+1)/2$ quarks, whose indices are saturated by $n+1$ epsilon tensors.
So what we have shown is that  at
least $n(n+1)$  quarks are necessary if we require the complete antisymmetry of the gauge wave function. This is equivalent to say that we put a lower-bound on the number of quarks: $S \geq n(n+1)/2$.

Now we need to prove the existence and uniqueness of this wave function. Consider the tensorial
product of a certain number of quarks $Q^{\{\alpha\beta\}}$. Every quark must be
considered as a vector space of dimension $n(n+1)/2$ over which the
group $\SU(n)$ acts as a linear representation. Now we take the subspace of the
tensor product that is completely antisymmetric under the exchange. This subspace
is obviously closed under the action of the gauge group. If the number of
quarks is greater than $n(n+1)/2$, this subspace has dimension zero, and this is equivalent to an upper-bound on the number of quarks: $S \leq n(n+1)/2$.
Due to the previously found lower-bound, we can say that $S$ must be exactly $n(n+1)/2$. 
If the number of quarks is exactly $n(n+1)/2$, the antisymmetric subspace
has exactly dimension one. We have thus proved that the completely antisymmetric space
of $n(n+1)/2$ quarks $Q^{\{\alpha\beta\}}$ is also a singlet of the
gauge group, since it is one-dimensional and must be closed under the gauge transformations.
\end{proof}

The gauge wave function for general $n$ can be obtained by generalizing the
one of Figure \ref{duesymmetric} for $n=2$.
The baryon for $n=2$ does not need to be
antisymmetrized, because it is already antisymmetric under the exchange of any pair
of quarks.  For generic $n$, antisymmetrization is needed. In the case $n=3$, for example, the antisymmetrizations with respect to the four quarks
$Q^{\{\alpha_{1}\beta_{1}\}}$, $Q^{\{\alpha_{2}\beta_{2}\}}$, $Q^{\{\alpha
_{3}\beta_{3}\}}$ and $Q^{\{\alpha_{4}\beta_{4}\}}$ are enough to guarantee the
complete antisymmetrization.   It can be seen that the antisymmetrization with respect to the exchange
$Q^{\{\alpha_{1}\beta_{1}\}}\leftrightarrow$ $Q^{\{\alpha_{2}\beta_{2}\}}$
implies that with respect to\ \ $Q^{\{\alpha_{3}\beta_{3}\}}\leftrightarrow
Q^{\{\alpha_{5}\beta_{5}\}}$ and the same for the two exchanges $Q^{\{\alpha
_{2}\beta_{2}\}}\leftrightarrow Q^{\{\alpha_{4}\beta_{4}\}}$ and
$Q^{\{\alpha_{3}\beta_{3}\}}\leftrightarrow Q^{\{\alpha_{6}\beta_{6}\}}$. We
 thus have a sufficient number of exchanges to generate the complete
permutation group.

\subsubsection{The antisymmetric representation}

We now consider the quarks in the antisymmetric representation. Our goal is a gauge invariant and antisymmetric
wave function that contains $n(n-1)/2$ quarks $Q^{[\alpha\beta]}$.
As we did in the previous subsection, we start with the simplest cases and
then generalize.  For $n=2$, we have $n(n-1)/2=1$, and it is easy to find such a wave
function. It is just $\epsilon_{\alpha\beta}\,Q^{[\alpha\beta]}$. For $n=3$, we need a
wave function that contains three quarks. To guess it directly from 
$Q^{[\alpha\beta]}$ is not easy, but we can use an indirect trick. The antisymmetric
representation, for $n=3$, is equivalent to the anti-fundamental representation; this follows from  $\widetilde
{Q}_{\gamma}=\frac{1}{2}\epsilon_{\gamma\alpha\beta}Q^{[\alpha\beta]}$.  We
know how to write a baryon for the anti-fundamental representation; it is the usual one $\epsilon^{\gamma\rho\tau}\widetilde{Q}_{\gamma}\widetilde{Q}_{\rho}%
\widetilde{Q}_{\tau}$.
Substituting the relationship between $\widetilde{Q}_{\gamma}$ and $Q^{[\alpha
\beta]}$, we obtain%
\begin{equation}
\frac{1}{2}(\epsilon_{\alpha_2\beta_2\alpha_1}\epsilon_{\alpha_3
\beta_3\beta_1}-\epsilon_{\alpha_3\beta_3\alpha_1}\epsilon_{\alpha_2\beta_2\beta_1})\,Q^{[\alpha_1\beta_1]}Q^{[\alpha_2\beta_2]}%
Q^{[\alpha_3\beta_3]}~. \label{treantisymmetricbaryon}%
\end{equation}
The diagram corresponding to this baryon is given in Figure \ref{treantisymmetricbaryon}.
\begin{figure}[th]
\begin{center}
\leavevmode \epsfxsize 2.2 cm \epsffile{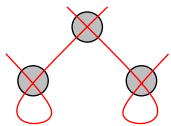}
\end{center}
\caption{\footnotesize  Diagrammatic representation for the simplest baryon of Eq.~(\ref{treantisymmetricbaryon}).}
\label{treantisymmetricbaryon}
\end{figure}
We know, by construction, that this wave function is antisymmetric under the
exchange of any couple of quarks. 
We now generalize this result.

\begin{proposition}
\label{proposition2}
\ There is one and only one gauge wave function that is a gauge singlet and
completely antisymmetric under the exchange of two quarks. This wave function is
composed by $n(n-1)/2$ quarks $Q^{[\alpha\beta]}$ and is the
antisymmetric subspace of the tensor product of $n(n-1)/2$ quarks
$Q^{[\alpha\beta]}$.
\end{proposition}
\begin{proof}
Denote by $A$ the number of quarks in a hypothetical gauge wave function that
satisfies the previous conditions. The reason why $A$ can be smaller than $S$
is that now it is instead possible for a quark to have both indices on the
same saturation line. For the proof we need the following two
basic facts: ${\it 1)}$ One saturation line can contain at most one quark; otherwise the wave function will be symmetric
under the exchange of these quarks; ${\it 2)}$ Two quarks $Q^{\{\alpha
_{i}\beta_{i}\}}$ and $Q^{\{\alpha_{j}\beta_{j}\}}$ can be connected by {\it at most}  one line. The reason is the
same as in the case of the symmetric representation. At this point, we are ready to build the fully antisymmetric wave function. We follow  Figure \ref{generalantisymmetric}.
\begin{figure}[th]
\begin{center}
\leavevmode \epsfxsize 6.5 cm \epsffile{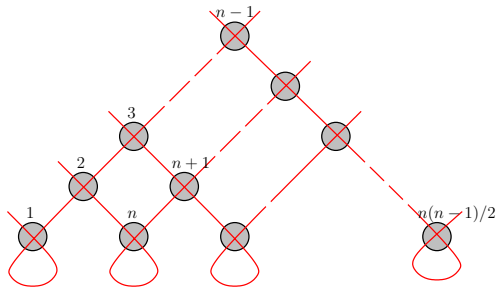}
\end{center}
\caption{{\protect\footnotesize Diagram for the Skyrmion-Baryon in the case of the two-index antisymmetric representation. }}
\label{generalantisymmetric}
\end{figure}
We start from the first quark $Q^{\{\alpha_{1}\beta_{1}\}}$ and draw the first saturation line that departs from this quark. This implies the presence of other $n-2$ quarks at least that we call $Q^{\{\alpha_{2}\beta_{2}\}} , \dots, Q^{\{\alpha_{n-1}\beta_{n-1}\}}$. 
Let us consider $Q^{\{\alpha_{2}\beta_{2}\}}$. One index is already saturated, and from the other one a new saturation line must depart.  Due to property ${\it 2)}$, at least new $n-2$ quarks must be added. One of them,  $Q^{\{\alpha_{n}\beta_{n}\}}$,  can have two indices on the same line. We then consider $Q^{\{\alpha_{3}\beta_{3}\}}$ and start the other saturation line. It can pass from $Q^{\{\alpha_{n+1}\beta_{n+1}\}}$, but then other different $n-3$ quarks must be added. So on we go until we reach $Q^{\{\alpha_{n-1}\beta_{n-1}\}}$ and complete the saturation adding the last line. In total, we have $n-1 + n-2 + \dots +1 = n(n-1)/2$ quarks, whose indices are saturated by $n-1$ epsilon tensors.
We just have proved that  at
least $n(n-1)$  quarks are necessary if we require the complete antisymmetry of the gauge wave function; this is the lower-bound $A \geq n(n-1)/2$.

The proof of the existence and uniqueness of this wave function is exactly the same as that of the symmetric representation. From the antisymmetric subspace of the tensorial product, we find the upper-bound $A \leq n(n-1)/2$  This implies that the unique possibility is exactly $A=n(n-1)/2$. From the fact that the antisymmetric subspace is, in this case, one-dimensional, follows the gauge invariance.
\end{proof}

For example, the baryon for $n=4$ is given by the diagram  plus the needed antisymmetrizations. 
\bea
&  \left(  \sum_{\sigma\in S} \mathrm{sign(\sigma)} \epsilon_{\alpha
_{\sigma(4)}\beta_{\sigma(4)}\alpha_{\sigma(2)}\alpha_{\sigma(1)}%
\epsilon_{\sigma(2)}\beta_{\sigma(5)}\beta_{\sigma(2)}\alpha_{\sigma(3)}%
}\epsilon_{\alpha_{\sigma(6)}\beta_{\sigma(6)}\beta_{\sigma(1)}\beta
_{\sigma(3)}} \right) \nonumber\\
&  Q^{\{\alpha_{1}\beta_{1}\}}Q^{\{\alpha_{2}\beta_{2}\}}Q^{\{\alpha_{3}%
\beta_{3}\}} Q^{\{\alpha_{4}\beta_{4}\}}Q^{\{\alpha_{5}\beta_{5}\}}%
Q^{\{\alpha_{6}\beta_{6}\}} \ .
\eea

\subsection{More on the Antisymmetric Representation} 
\label{moreantisymmetric}

In this section, we want to consider in more detail the case of the
antisymmetric representation. There is a peculiarity about the simplest baryon, which we did not mention previously.
The baryon previously introduced as the simplest one, is that of Figure \ref{basic}, which is
\begin{equation}
\epsilon_{\alpha_{1}\alpha_{2}\dots\alpha_{n}}\epsilon_{\beta_{1}\beta
_{2}\dots\beta_{n}}\,Q^{[\alpha_{1}\beta_{1}]}Q^{[\alpha_{2}\beta_{2}]}\dots
Q^{[\alpha_{n}\beta_{n}]}~ \ . \label{minimal}%
\end{equation}

We have now to make a distinction between $n$ even and $n$ odd. In the case of
$n$ even, (\ref{minimal}) is not the minimal baryon, since we can construct a
gauge invariant wave function using only $n/2$ quarks:%
\begin{equation}
\epsilon_{\alpha_{1}\alpha_{2}\dots\alpha_{n/2}\beta_{1}\beta_{2}\dots
\beta_{n/2}}\,Q^{[\alpha_{1}\beta_{1}]}Q^{[\alpha_{2}\beta_{2}]}\dots
Q^{[\alpha_{n/2}\beta_{n/2}]}~.
\end{equation}
This baryon is symmetric under the exchange of two quark and so there is no
difference with respect to the previous one with regard to the mass-vs-$n$ dependence.

In the case of $n=2n+1$, instead we can prove that the minimal
baryon (\ref{minimal}) is identically zero, with the following algebraic
passages:%
\bea
&  ~~~~~~~~~~~~~~~~~\epsilon_{\alpha_{1}\alpha_{2}\dots\alpha_{2n+1}}%
\epsilon_{\beta_{1}\beta_{2}\dots\beta_{2n+1}}\,Q^{[\alpha_{1}\beta_{1}%
]}Q^{[\alpha_{2}\beta_{2}]}\dots Q^{[\alpha_{2n+1}\beta_{2n+1}]}\nonumber\\
&  =\,\left(  -1\right)  ^{2n+1}\epsilon_{\alpha_{1}\alpha_{2}\dots
\alpha_{2n+1}}\epsilon_{\beta_{1}\beta_{2}\dots\beta_{2n+1}}\,Q^{[\beta
_{1}\alpha_{1}]}Q^{[\beta_{2}\alpha_{2}]}\dots Q^{[\beta_{2n+1}\alpha_{2n+1}%
]}\nonumber\\
&  =~~~~~~~~~-\epsilon_{\beta_{1}\beta_{2}\dots\beta_{2n+1}}\epsilon
_{\alpha_{1}\alpha_{2}\dots\alpha_{2n+1}}Q^{[\alpha_{1}\beta_{1}]}%
Q^{[\alpha_{2}\beta_{2}]}\dots Q^{[\alpha_{2n+1}\beta_{2n+1}]}~.
\label{passages}%
\eea
In the first passage, we have exchanged the $\alpha$ and the $\beta$ indices in
every quark. Since we have $2n+1$ quarks in the antisymmetric representation,
this step brings down a minus sign. In the second step we have just renamed
$\alpha_{i}$ with $\beta_{i}$ and vice versa, and this has no consequences. The
last line of (\ref{passages}) is equal to minus the first line (apart from an
irrelevant exchange in the position of the two epsilons), and thus the wave
function must be zero. We can also prove a stronger statement:

\begin{proposition}
For $n$ odd and quarks in the antisymmetric representation, it is not possible
to write a gauge invariant wave function that is completely symmetric under the
exchange of two quarks.
\end{proposition}

\begin{proof}
Consider a generic wave function that is gauge invariant and symmetric under the
exchange of two quarks. We are going to prove that it is identically zero.
This wave function is composed by a number of quarks that we generically
denote by $M$. $M_{\alpha\beta}$ of these quarks are of type $Q^{[\alpha
\beta]}$ and $M_{\alpha\beta}$ are of type $Q^{[\alpha\beta]}$ so that we
can write%
\begin{equation}
M=M_{\left(  \alpha\beta\right)  }+M_{\left(  \alpha\beta\right)  }~.
\label{emme}%
\end{equation}
The $M$ quarks can be divided into various \emph{connected} components, where
the connection is given by the epsilon contractions and the quarks
$Q^{[\alpha\beta]}$. Let us assume for the moment that we have only one
connected component. It is easy to see that $M_{\alpha\beta}$ must be odd. We
will now\ use the same argument we have used to show that (\ref{passages}) is
identically zero. Namely we will show that the wave function is equal to minus
itself. First we exchange all the $\alpha$ indices with their $\beta$ partners
and this contributes a minus sign since $M_{\alpha\beta}$ is odd. Then we make
a suitable number of exchanges between the quarks $Q^{[\alpha\beta]}$ in order
to recover the original epsilon structure. These exchanges do not affect the
wave function since by definition it is symmetric under exchanges of two
quarks. So we have recovered the original wave function but with a minus sign
in front.

We now have to consider the more general situation in which the $M$ quarks are
divided into various disconnected components. It can easily be seen that in this
case the sub-connected components must be closed under the exchange of two
generic quarks. Put in another way, if the global wave function is symmetric
under the exchange of two quarks, then the sub-connected wave functions are also
symmetric under the exchange of two quarks.
\end{proof}

The previous proposition does not exclude the possible existence of a gauge
invariant wave function with less than $n(n-1)/2$ quarks, and in a non-singlet representation of the permutation group. In this case, the baryon is
not a simple product of gauge, spin, and space wave function but a sum
$\sum_{i}\psi_{\mathrm{gauge}}^{i}\psi_{\mathrm{spin}}^{i}\psi_{\mathrm{space}%
}^{i}$, where $\psi_{\mathrm{gauge}}^{i}$ is the non-singlet representation of
the permutation group.

\subsection{Stability of the Skyrmion}
\label{stabilitytwo}

We want now to discuss the issue of the stability of the Skyrmion. The
Skyrmion corresponds to the baryon that contains $n(n\pm 1)/2$ quarks and is fully antisymmetric in the gauge wave function. The
mass is thus proportional to the number of constituent quarks. Seen from the
low-energy effective Lagrangian, the Skyrmion is absolutely stable. In the full
theory, on the other hand, we should consider the possibility of decay into
baryons with lower numbers of constituent quarks, for example, the baryon
$\epsilon_{\alpha_{1}\alpha_{2}\dots\alpha_{n}}\epsilon_{\beta_{1}\beta
_{2}\dots\beta_{n}}\,Q^{\alpha_{1}\beta_{1}}Q^{\alpha_{2}\beta_{2}}\dots
Q^{\alpha_{n}\beta_{n}}$. These states are not visible from the low-energy
effective Lagrangian. As we have seen in Section \ref{stablebaryons}, baryons
with a number of constituent quarks lower than $n(n\pm 1)/2$ cannot be in a fully antisymmetric gauge wave function. This implies that
the Skyrmion is the state that minimizes the mass per unit of baryon number.

\begin{figure}[h]
\begin{center}
\leavevmode \epsfxsize 6.5 cm \epsffile{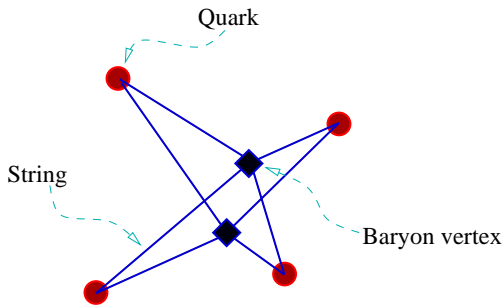}
\end{center}
\caption{{\protect\footnotesize A model of the baryon (here for four colors).
Every quark is attached to two confining strings and every baryon vertex to
$n$ confining strings.}}
\label{toymodel}
\end{figure}

\begin{figure}[h]
\begin{center}
\leavevmode \epsfxsize 8 cm \epsffile{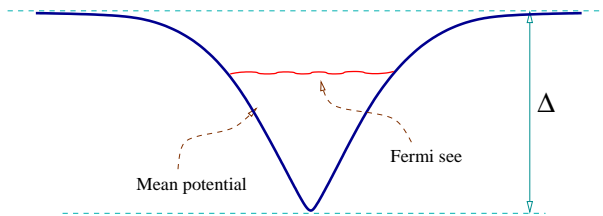}
\end{center}
\caption{{\protect\footnotesize The mean potential for our toy model of the baryon.}}
\label{potentialbaryon}
\end{figure}

Let us consider an explicit example in more detail. A Skyrmion that contains
$n(n\pm 1)/2$ can decay into $(n\pm 1)/2$ baryons
composed by $n$ quarks. The baryon number is conserved, and so this decay
channel is in principle possible. In order to analyze the energetic of this
baryon, we propose now a toy model to schematize the fundamental baryon. We
have $n$ quarks and $2$ baryon vertices. Every quark is attached to two
fundamental strings and every baryon vertex to $n$ fundamental strings (see
Figure \ref{toymodel} for an example).
Baryon vertices have a mass of order $n$; we can thus neglect their dynamics
and consider them at rest and positioned in what we define to be the center of
the baryon. In this approximation, the quarks do not interact directly between
each other; they live in a mean potential given by the string tension
multiplied by the distance from the center%
\begin{equation}
V_{\mathrm{mean}}(R)=2T_{\mathrm{string}}\left\vert R\right\vert
~.\label{confiningpotential}%
\end{equation}
Quarks are antisymmetric in the space wave function, and so they fill the
energy levels up to the Fermi surface (see Figure \ref{potentialbaryon}). We
indicate as $R_{\mathrm{F}}$ and $P_{\mathrm{F}}$, respectively, the Fermi
radius and momentum. The total energy and the number of quarks $n$ are given
by the following integrals over the phase space:%
\bea
\int^{R_{\mathrm{F}}}\int^{P_{\mathrm{F}}}\frac{d^{3}Rd^{3}P}{\left(
2\pi\right)  ^{3}}\left(  P+V_{\mathrm{mean}}(R)\right)   &  =E~,\nonumber\\
\int^{R_{\mathrm{F}}}\int^{P_{\mathrm{F}}}\frac{d^{3}Rd^{3}P}{\left(
2\pi\right)  ^{3}} &  =n~.\label{integrals}%
\eea
Since the quarks are massless, we take the Hamiltonian to be
$P+V_{\mathrm{mean}}(R)$. From now on, we ignore numerical factors such as the
phase space volume element; at this level of approximation they are not
important. The second equation of (\ref{integrals}) gives a relationship between
the Fermi momentum and the Fermi radius, namely $P_{\mathrm{F}}\sim
n^{1/3}/R_{\mathrm{F}}$. The first equation of (\ref{integrals}) gives the
following expression of the energy as function of the radius
\begin{equation}
E\sim\frac{n^{4/3}}{R_{\mathrm{F}}}+T_{\mathrm{string}}nR_{\mathrm{F}}~.
\end{equation}
Minimizing, we obtain $R_{\mathrm{F}}\sim n^{1/6}/\sqrt{T_{\mathrm{string}}}%
$,$\ $and\ consequently$\ P_{\mathrm{F}}\sim n^{1/6}\sqrt{T_{\mathrm{string}}%
}$. The mass of the baryon is thus given by%
\begin{equation}
M_{n-\mathrm{Baryon}}\sim n^{7/6}\sqrt{T_{\mathrm{string}}}~.
\end{equation}
The important thing to note is the $n^{7/6}$ dependence. The mass per unit of
baryon number grows as $n^{1/6}$.

This approximation breaks down when the Fermi energy $n^{1/6}\sqrt
{T_{\mathrm{string}}}$ becomes much greater than the dynamical scale. Due to
asymptotic freedom, the highly energetic quarks do not feel a confining
potential like (\ref{confiningpotential}) but instead a Coulomb-like
potential. The mass per unit of baryon number stops growing as $n^{1/6}$ and, presumably,
saturates to a constant.

This toy model is certainly a very crude approximation. But we think it captures some qualitative behavior of this simplest baryon. 
In particular, the model shows how this Fermi statistic is responsible for the stability of the Skyrmion. 
The simplest baryon, in the very large-$n$ limit, consists of a core of quarks in a confined potential, plus a cloud of quarks in a Coulomb-like potential. The mass per unit of baryon number exceeds that of the Skyrmion a the quantity $\Delta$ (in Figure \ref{potentialbaryon}), due to the Fermi statistic of the quarks. The Skyrmion is thus stable against decay into $\sim n$ of these simplest baryons.

Regarding the mass-vs.-$n$ dependence of the Skyrmion, there is a final issue we have to discuss now. The Skyrmion has $n(n \pm 1)/2$ constituent quarks, symmetric in the space wave function. The mass, as we said, is proportional to $n^2$ since all the quarks can occupy the ground state of the mean potential. But we should also consider higher order corrections, from the gluons exchange, and verify that they all scale like $n^2$. In ordinary QCD, that is indeed the case. A two-body interaction has a $\sim n^2$ enhancement due to the possibility of choosing any couple out of the $n$ quarks. On the other hand, there is also a $1/n$ suppression from the gauge coupling that enters in the gluon exchange. At any order, the corrections always scale like $n$.  If we simply repeat the same argument for the orientifold QCD, we run into a problem. Since we have $\sim n^2$ constituent quarks, the combinatorial enhancement factor is now $\sim n^4$. We still have a suppression of $1/n$ from the gauge coupling, and as a result the one-body interaction seems to grow as $\sim n^3$.  And things gets even worse if we consider higher order interactions. Needless to say, this is a problem.  Our result, the identification of the Skyrmion with the $n(n \pm 1)/2$ baryon, deeply relies on the fact that this object has a mass that scales like $\sim n^2$.

This issue has been considered, and successfully solved, in Ref.~\cite{Cherman:2006iy}. The point is that, in the previous paragraph, we overestimated the combinatorial factor. The gauge structure of the baryon forbids many gluon exchanges between quarks, reducing the combinatorial factor from order $n^4$ to order $n^3$. Suppose we want to exchange a gluon between quark $Q^{12}$ and quark $Q^{34}$, where the numbers refer to the gauge space. We can do it with the gluon $A_{\mu}^{23}$, for example. The outcome is that the two quarks exchange the gauge numbers carried by the gluon. We thus have that the quarks $Q^{12}$ and $Q^{34}$ become $Q^{13}$ and $Q^{24}$. But the baryon already contains quarks  $Q^{13}$ and $Q^{24}$ in its wave function, and the completely antisymmetric structure forbids  repetitions. That means that this gluon exchange is not allowed. The only exchanges allowed  are the ones between quarks that share at least one index. Quarks $Q^{\alpha \beta}$ and $Q^{\alpha \gamma}$ can interact exchanging the gluon $A_{\mu}^{\beta \gamma}$ or the diagonal one $A_{\mu}^{\alpha \alpha}$. This reduces the total combinatorial factor from $n^4$  to $n^3$. This, together with the $1/n$ suppression from the gauge coupling, gives a contribution of order $n^2$, which is exactly what we expect from the Skyrmion-baryon identification. Repeating the same argument, for higher body interactions, still gives a total result of order $\sim n^2$.

\section{Adjoint QCD}
\label{adjoint}
We will now consider adjoint QCD: $n_f $ massless Weyl fermions in the adjoint representation of the SU$(n)$ gauge
theory.  To ensure a chiral symmetry in the fundamental Lagrangian, and to keep the microscopic theory asymptotically free, we must impose the following constraints on the value of $n_f$:
\begin{equation}
2\leq n_f\leq 5\,.  \label{fc}
\end{equation}
The global symmetries are $\U(n_f)$ group acting on the quarks $\lambda_f$.

The pattern of the $\chi$SB in this case is
\begin{equation}
\mathrm{SU}(n_f)\times \mathbb{Z}_{2n n_f}\to \mathrm{SO}(n_f)
\times \mathbb{
Z} _{2} \,  ,  \label{pater}
\end{equation}
where the discrete factors are the remnants of the global $\U(1)$ broken by the anomaly. 
Equation (\ref{pater}) can be elucidated as follows. 
If we denote the
adjoint quark field as $\lambda^a_{\alpha\, f}$ (here $a,\,\alpha ,\, f$ are
the color, Lorentz-spinorial, and flavor indices, respectively; we use the
Weyl representation for the spinor), the Lorentz-scalar bilinear $%
\lambda^a_{\alpha\, f}\, \lambda^{a\,\alpha}_{g}$ is expected to condense as:
\begin{equation}
\langle\lambda^a_{\alpha\, f}\, \lambda^{a\,\alpha}_{g} \rangle =\Lambda^3\,
\delta_{fg} \,.
\end{equation}
The order parameter $\langle\lambda^a_{\alpha\, f}\, \lambda^{a\,\alpha}_{f}
\rangle$ stays intact under the transformations of SU$(n_f)$ that are
generated by the purely imaginary generators, the ones of the so$(n_f)$ sub-algebra.

Thus, the low-energy pion Lagrangian is a nonlinear sigma model with the
target space $\mathcal{M}$ given by the coset space:
\begin{equation}
\mathcal{M}_{n_f} = \mathrm{SU}(n_{f})/\mathrm{SO}(n_{f})\,.  \label{target}
\end{equation}
The third
homotopy group for (\ref{target}) is nontrivial in all four cases (\ref{fc}),
as shown in Table~\ref{tabone}.
\begin{table}[htb]
\begin{center}
\begin{tabular}{|c|c|c|c|c|}
\hline
\rule{0mm}{5mm} & $n_{f}=2$ & $n_{f}=3$ & $n_{f}=4$ & $n_{f}=5$ \\
[1mm] \hline
\rule{0mm}{5mm} $\pi_{3}$ & $\mathbb{Z}$ & $\mathbb{Z}_{4}$ &
$\mathbb{Z} _{2} $ & $\mathbb{Z} _{2}$ \\[1mm] \hline
\end{tabular}
\end{center}
\caption{{\protect\footnotesize The third homotopy group for sigma models
emerging in Yang--Mills with two, three, four, and five adjoint flavors.}}
\label{tabone}
\end{table}

Topologically stable solitons (whose mass scales as $n^2$)
exist much in the same way as Skyrmions in QCD. 
Unlike QCD, where the
relationship between Skyrmions and microscopic theory is well established, in
our case it is far from being clear.

Below we will clarify this aspect of the theory. We will first focus on
the simplest case $n_f=2$ in Section \ref{due}. In this case, the chiral
Lagrangian is that of the $S^2$ sigma model,  which is sometimes referred
to as the Skyrme-Faddeev and sometimes as the Faddeev-Hopf model.
The soliton's topological stability is due to the existence of the Hopf
invariant. We will then  discuss the generic case  $n_f =3,4,5$ in Section \ref{general} and the related SO$(n)$ Yang--Mills theories with
vectorial quarks, which share the same structure of chiral Lagrangians and solitons.

\subsection{The Case of Two Flavors}
\label{due}

In the $n_f=2$ case, the SU$(2)$ flavor group is broken down to the $\U(1)$ subgroup generated by
the Pauli matrix $\tau_{2}$ (we will denote the Pauli matrices as $\tau_{i}$
when they act on the flavor indices and $\sigma_{i}$ when they act in the
Lorentz-spinorial indices.) 
If $\lambda\to \exp (i\alpha\tau_2)\lambda$, we
can diagonalize the adjoint quarks in the flavor space as follows:
\begin{equation}
\lambda_{\pm} =\frac{1}{\sqrt{2}}\left( \lambda_{1}\mp i\lambda_{2}\right)
\,.
\end{equation}
Then, the unbroken U$(1)$ current takes the form
\begin{equation}
J_{\alpha\dot\alpha} = \bar\lambda_{+\dot\alpha}\, \lambda_{+\alpha} -
\bar\lambda_{-\dot\alpha}\, \lambda_{-\alpha} \ ,
\end{equation}
generating the following transformation:
\begin{equation}
\lambda_+\to e^{i\alpha }\lambda_+\,,\qquad \lambda_-\to e^{-i\alpha
}\lambda_-\,.
\end{equation}
We will say that the U$(1)$ charge of $\lambda_+$ is plus one while that of $%
\lambda_-$ is minus one, $Q(\lambda_{\pm})=\pm1$. The vacuum condensate in
this theory has the form
\begin{equation}
\langle \lambda_+\lambda_- +\mbox{h.c.}\rangle \neq 0\,.
\end{equation}
It is neutral with respect to the conserved U(1). There are two
Nambu--Goldstone bosons, $\pi^{++}$ and $\pi^{--}$. Roughly, $\pi^{++}\sim
\lambda_+\lambda_+$ and $\pi^{--}\sim \lambda_-\lambda_-$. 
The pion U(1) charges
are
\begin{equation}
Q(\pi^{\pm\pm} ) = \pm 2\,.
\end{equation}

For all ``ordinary" hadrons, which can be produced from the vacuum by local
currents,  determination of $(-1)^F$ is
straightforward. 
We can decompose the Hilbert space of hadronic excitations in the
direct sum of two spaces:
\begin{equation}
\mathcal{H}^{\left(\mathrm{hadronic}\right)}=\mathcal{H}^{\left(+1,+1\right)
}\oplus\mathcal{H} ^{\left(-1,-1\right) } \label{Hilbert senza
fractional}
\end{equation}
containing, respectively, the composite states with the even and odd
U(1) charges. 
We have denoted the charges as
$\left((-1)^{Q},(-1)^{F}\right)$. 
From the point of view of the
hadronic Hilbert space (\ref{Hilbert senza fractional}), this would
appear a redundant notation. It will soon be clear that this is not.
It is clear that $\mathcal{H}^{\left(+1,+1\right)
} $
contains hadronic excitation of the boson type, while $\mathcal{H} ^{\left(-1,-1\right) }$ of the fermion type. 
In particular, $\mathcal{H}^{\left(+1,+1\right)
}$ contains the massless Nambu-Goldstone
bosons $\pi^{\pm\pm}$, and, hence, there is no mass gap here. 
On the
contrary, $\mathcal{H} ^{\left(-1,-1\right) }$ has a mass gap $m$,
the mass of the lightest composite fermion of the type
\begin{equation}
\psi_{\beta\,\,f} \propto \mathrm{Tr}\left( \lambda^{\alpha}_f\,F_{\alpha{\beta}%
}\right) \equiv \mathrm{Tr}\left( \lambda^{\alpha}_f \sigma_{\alpha{\beta}%
}^{\mu\nu}F_{\mu\nu}\right),  \label{mafe}
\end{equation}
where $F_{\alpha\beta}$ is the (anti)self-dual gluon field strength tensor
(in the spinorial notation). 
Two U(1)-charge $\pm 1$ composite fermions are
\begin{equation}
\psi_\pm  \propto \mathrm{Tr}\left( \lambda_\pm^{\alpha}F_{\alpha{\beta}}\right) \,,
\end{equation}
(plus their antiparticles, of course). Note that $\psi_-$ is \emph{not} $%
\psi_+$'s antiparticle. Moreover, we can combine $\psi_{-}$ and $%
\bar\psi_{+} $ in a single Dirac spinor $\Psi_{-}$,
\begin{equation}
\Psi_{-} =\left(
\begin{array}{c}
\psi_{-} \\[1mm]
\,\sigma_2 \, \tau_2\, {\psi}_{+}^*%
\end{array}
\right) \,.
\end{equation}
\mbox{} \vspace{1mm}

Below we will argue that in fact Eq.~(\ref{Hilbert senza fractional}) is
incomplete. An extra sector can and must be added,
\begin{equation}
\mathcal{H}=\mathcal{H}^{\left(\mathrm{hadronic}\right)}
\oplus\mathcal{H}^{\left( \mathrm{exotic} \right) } \,,
\label{extras}
\end{equation}
where $\mathcal{H}^{\left(\mathrm{hadronic}\right)}$ is given by
(\ref{Hilbert senza fractional}) and the new sector is given by
\begin{equation}
\mathcal{H}^{\left(\mathrm{exotic}\right)}=\mathcal{H}^{\left(+1,-1\right)
}\oplus\mathcal{H} ^{\left(-1,+1\right) }\ .
\end{equation}
$\mathcal{H}^{\left( \mathrm{exotic}\right) }$ includes hadrons with
even $Q$ and odd $F$ and, vice versa, odd $Q$ and even $F$. To build
such a hadron, one needs $\sim n^2$ constituents. In adjoint QCD, they
play the role of baryons of conventional QCD. Their existence is
reflected in the Hopf-Skyrmions.

The sections are organized as follows. In \ref{skyrme}, we discuss the low-energy limit of two-flavor
adjoint QCD, present the corresponding chiral Lagrangian, and discuss its
features.   In \ref{massive}, we extend the model by introducing
appropriate fermion fields. In \ref{impa}, we calculate the induced
fermion charge, and thus prove the stability.

\subsubsection{The Skyrme--Faddeev model}
\label{skyrme}

Let us briefly review the effective low-energy pion Lagrangian
corresponding to the given pattern of the $\chi$SB; see Eq. (\ref{pater})
with $n_f=2$. We describe the pion dynamics by the $O(3)$ nonlinear sigma
model (in four dimensions)
\begin{equation}
\mathcal{L}_{\mathrm{eff}}=\frac{F_{\pi}^{2}}{2}\,\, \partial_{\mu}\vec{n}
\cdot\partial^{\mu}\vec{n} +\mbox{higher derivatives} \ ,
\label{senza il fermione}
\end{equation}
where the three-component field $\vec n$ is a vector in the \emph{flavor}
space, subject to the condition
\begin{equation}
\vec n^{\,2} =1\ .  \label{ts}
\end{equation}
As in ordinary QCD,   higher derivative terms are in general present in a low-energy effective theory expansion, and they are needed for the soliton stabilization. 
The ``plain"
vacuum corresponds to a constant value of $\vec n$, which we are free to
choose as $\langle n_3\rangle =1$.

Usually the higher derivative term is chosen as follows (for a review, see
\cite{manton}):
\begin{equation}
\delta \mathcal{L}_{\mathrm{eff}} = -\frac{\lambda}{4}\,
\left(\partial_{\mu} \vec{n}\times\partial_{\nu}\vec{n} \right)
\cdot\left(\partial^{\mu}\vec{n} \times\partial^{\nu}\vec{n} \right)\,.
\label{senza il fermionep}
\end{equation}
Equations (\ref{senza il fermione}) and (\ref{senza il fermionep})
constitute the Skyrme--Faddeev (or the Faddeev--Hopf) model. Note that the
WZNW term does not exist in this
model, simply because $\pi_4(S^2)$ is not trivial.

To have a finite soliton energy, the vector $\vec n$ for the soliton solution
must tend to its vacuum value at the spatial infinity,
\begin{equation}
\vec n\to \{0,0,1\}\,\,\,\mbox{at}\,\,\, \left| \vec x \right| \to\infty\,.
\label{vac}
\end{equation}
Two elementary excitations near the vacuum $n_3 =1$,
\[
\frac{1}{\sqrt 2}\left(n_1\pm i\, n_2\right),
\]
can be identified with the pions. 
The boundary condition (\ref{vac})
compactifies the space to $S^3$. Since $\pi_3 (S^2) = \mathbb{Z}$, see Table~
\ref{tabone}, solitons present topologically nontrivial maps of $S^3\to S^2$
. As was noted in \cite{Faddeev}, there is an associated integer topological
charge $n$, the Hopf invariant, which presents the soliton number.

\begin{figure}[th]
\begin{center}
\leavevmode \epsfxsize 8.5 cm \epsffile{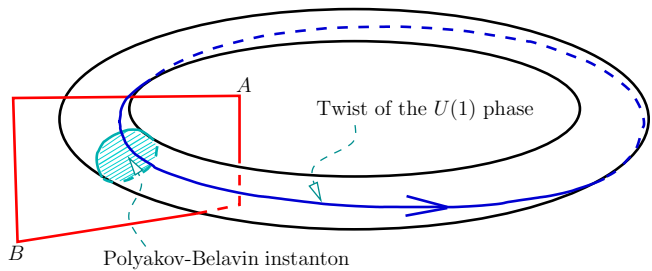}
\end{center}
\caption{{\protect\footnotesize The simplest Hopf soliton, in the adiabatic
limit, corresponds to a Belavin-Polyakov soliton closed into a donut after a $
2\protect\pi$ twist of the internal phase.}}
\label{donu}
\end{figure}

The solitons of the Skyrme--Faddeev model are of the knot type. The
simplest of them is toroidal; it looks like a ``donut," see, e.g., \cite{manton}. 
Qualitatively, it is rather easy to understand, in the limit when
the ratio of the periods is a large number, that the Hopf
topological number combines an instanton number in two dimensions,
with a twist in the perpendicular dimension. Let us slice the ``donut"
soliton by a perpendicular plane $AB$. In the vicinity of this plane,
the soliton can be viewed as a cylinder, so that the problem becomes
effectively two-dimensional. In two dimensions the, O(3) sigma model
has Polyakov--Belavin instantons \cite{Polyakov} whose topological
stability is ensured by the existence of the corresponding
topological charge. The Polyakov--Belavin instanton has an
orientational collective coordinate describing its rotation in the
unbroken U(1) subgroup. In two dimensions for each given instanton,
this collective coordinate is a fixed number. In the Hopf soliton of
the type shown in Fig.~\ref{donu}, as we move the plane $AB$ in the
direction indicated by the arrow, this collective coordinate changes
(adiabatically), so that the $2\pi$ rotation of the plane $AB$ in
the direction of the arrow corresponds to the $2\pi$ rotation of the
orientational modulus of the Polyakov--Belavin instanton. This is
the twist necessary to make the Hopf soliton topologically stable.

The Hopf charge {\it cannot} be written as an integral of any density
that is local in the field $\vec n$.
We introduce now a different, but equivalent, formulation.
In the $n_f=2$  case, there are two ways to parametrize the target space $S^2$.
One can use a vector $\vec{n}$ subject to the constraint $|\vec{n}|  =1$, the one just discussed.
 Another approach, which goes under the name of the
 gauged formulation of the $CP^1$ sigma model,  is to use a complex doublet $z_i$ subject to the constraint $z_i^* z_i =1$. This leaves us with an $S^3$ sphere. We have to further reduce it by gauging the phase rotation $z_i \to e^{i \theta} z_i$. This Hopf fibration leaves us exactly with the sphere $S^2$.  The map between the two formulations is
 $$
 \vec{n} = z^*_i \vec{\tau}  z_i\,.
 $$
 The derivatives  acting on the doublet $z_i$ are the covariant derivatives
 $$D_{\mu} = \partial_{\mu} -iA_{\mu}$$ where
\beq
A_{\mu} = -\frac{i}{2}\left[ z_i^*
({\partial}_{\mu}z_i)-({\partial}_{\mu}z_i^*) z_i\right] .
\eeq
This gauge field is related to the topological current of the vortex  by
\bea
J^{\mu} &=& \frac14 \epsilon^{\mu \nu \rho} \epsilon^{abc} n^a \partial_{\nu} n^b \partial_{\rho} n^c \\
 &=& \epsilon^{\mu \nu \rho} \partial_{\nu} A_{\rho} \ ,
\eea
where the normalization is to $2\pi$.

It is possible now to express the Hopf charge (the charge of  $\pi_3(S^2)=Z$) as a local function of the gauge field $A$, simply using the Chern-Simons term for the auxiliary gauge field:
\beq
s=\frac{1}{4\pi^2}  \int d^3x \epsilon^{\mu\nu\rho} A_{\mu} \partial_{\nu} A_{\rho}\,.
\label{hopfnumb1}
\eeq
We can verify this formula in the following way.
Since the Hopf charge is a topological invariant, we can compute it on a configuration with cylindrical symmetry, a vortex in the $x,y$ plane, extended in the $z$ direction, that makes a $2\pi$ twist of the internal phase, as $z$ goes from $-\infty$ to $+\infty$.  
Any configuration in this topological class  must give the same answer for the integral of the Hopf number. We can thus choose an adiabatic rotation, in which the $z$ variation happens at a scale much longer than the vortex one. This allows the following simplifications. We can neglect the term $A_i J_i$ and consider only the one $A_0 J_0$. Furthermore, for the same reason, we can make separately the integrals $\int dz A_0 \int d^2 x J_0$. The integral of $J_0$ gives the total magnetic flux, which is $2\pi$. The integral of $A_0$ gives the total phase rotation, which is also $2\pi$. From that, the $4\pi^2$ normalization factor in (\ref{hopfnumb1}).



\subsubsection{Introducing massive fermions}
\label{massive}

The Skyrme--Faddeev model neglects excitations belonging to the odd sector of
the Hilbert space. Let us now include them.

We can write an effective Lagrangian that includes both the pions and the
fermions as follows:
\begin{equation}
\mathcal{L}_{\mathrm{eff}}=F_{\pi }^{2}\left[ \frac{1}{2}\partial _{\mu }%
\vec{n}\cdot \partial ^{\mu }\vec{n}+\bar{\Psi}\left( i\gamma ^{\mu
}\partial _{\mu }-g\vec{n}\cdot \vec{\tau}\right) \Psi +\dots \right]
\label{lagrangiana}
\end{equation}%
where $\Psi $ is a Majorana spinor defined as
\begin{equation}
\Psi =\frac{1}{2F_{\pi }}\left(
\begin{array}{c}
\psi  \\
\sigma _{2}\tau _{2}\psi ^{\ast }%
\end{array}%
\right) \ .   \label{Majorana}
\end{equation}%
If we expand around $\vec{n}$ rotated in the third direction, we find as
expected the two pions plus a massive Dirac fermion:
\begin{equation}
\mathcal{L}_{\mathrm{eff}}= \partial _{\mu }\pi ^{*}\partial
^{\mu }\pi  + \bar{\Psi}_{\mathrm{D}}\left( i\gamma ^{\mu }\partial _{\mu }-m\right)
\Psi _{\mathrm{D}}+\mathrm{interactions~}  \ , \label{lagrangianslowly}
\end{equation}%
where the mass is $m=gF_{\pi }$.

\subsubsection{Fermion impact on the Hopf soliton}
\label{impa}

In a slow field configuration background, that is a
sufficiently wide soliton, the induced fermion quantum numbers can be
evaluated using the Goldstone-Wilczek(GW) technique \cite{Goldstone:1981kk}.

Now return  to our problem in $3+1$ dimensions. Now the Hopf
term
becomes a current%
\begin{equation}
j_{\mathrm{Hopf}}^{\mu }=\frac{1}{4\pi ^{2}}\epsilon ^{\mu \nu \rho \sigma
}A_{\nu }^{3}\partial _{\rho }A_{\sigma }^{3} \ ,
\end{equation}%
and it is normalized so that $\int d^{x}j_{\mathrm{Hopf}}^{0}=1$ on
the background of a Hopf Skyrmion of charge $1$. 
For a slow field
configuration, we can use the Goldstone-Wiczek method. We orient
$\vec{n}$ in the third direction and we obtain the Lagrangian
(\ref{lagrangianslowly}), that is one massive Dirac fermion $\Psi
_{\mathrm{D}}$ plus pions.
 The currents of the fermion are $j^{\mu
}=\bar{\Psi}_{\mathrm{D}}\gamma ^{\mu }\Psi _{\mathrm{D}}$ and \ \ \
$j_{5}^{\mu }=\bar{\Psi}_{\mathrm{D}}\gamma ^{\mu }\gamma ^{5}\Psi
_{\mathrm{D}}$ and the respective charges\ $Q=\int d^{3}xj^{0}$ and\
$Q_{5}=\int d^{3}xj_{5}^{0}$. $Q$ corresponds to the exactly
conserved $U(1)$ charge, while $Q_{5}$ is the conserved, modulus $2$,
fermionic number. The induced $Q$ charge is zero simply because the
left
fermion and the right fermion give opposite contributions. 
The induced $Q_{5}$ is not zero. 
We do need to compute this diagram,
since this has already been done: it \ is nothing but the ABJ
anomaly. The anomaly of
the axial current is%
\begin{equation}
\partial _{\mu }j_{5}^{\mu }=\frac{1}{8\pi ^{2}}\widetilde{F}^{\mu \nu
}F_{\mu \nu }+\mathrm{mass~,} 
\end{equation}%
which can just be rewritten as%
\begin{equation}
\partial _{\mu }j_{5}^{\mu }=\partial _{\mu }j_{\mathrm{Hopf}}^{\mu }+%
\mathrm{mass~.} 
\end{equation}%
The induced $Q_{5}$ charge is equal to the Hopf charge, modulus $2$.

We are now ready to discuss the stability of the Skyrmion. 
The theory contains three kinds of particle, whose charges are resumed in the Table \ref{tabtwo}.%
\begin{table}[tbp]
\begin{center}
\begin{tabular}{|c|c|c|}
\hline
& \textrm{Charge }$Q$ & $F $\rule{0mm}{5mm} \\ \hline
$\psi $ & $1$ & $1$\rule{0mm}{5mm} \\ \hline
$\pi $ & $2$ & $0$\rule{0mm}{5mm} \\ \hline
\textrm{Skyrmion/exotics} & {$
\begin{array}{c}
0 \\
1
\end{array}
$} & {$
\begin{array}{c}
1 \\
0
\end{array}
$} \rule{0mm}{8mm} \\ \hline
\end{tabular}
\end{center}
\caption{{\protect\footnotesize $Q$ and $F$ mod 2 for nonexotic and exotic
hadrons.}}
\label{tabtwo}
\end{table}
The Hopf Skyrmion can have charges $Q$ and $F$ respectively $0$ and $1$, if
there is no fermion zero mode crossing in the process of evolution from the
topologically trivial background to that of the Hopf Skyrmion, or $1$ and $0$
if there is a fermion zero mode crossing.\footnote{
A relevant discussion of the fermion zero mode crossings in 2+1 dimensions
can be found in \cite{Carena:1990vy}.}

In both cases the lightest exotic hadrons represented by the Hopf Skyrmion
are stable. They cannot decay in any number of pions and/or pions plus
``ordinary" baryons with mass $O(n^0)$. note that this is a $Z_{2}$
stability. Two Hopf Skyrmions can annihilate and decay into an array of $\pi$
's and $\psi$'s. For nonexotic hadron excitations that can be seen in a
constituent model and have mass $O(n^0)$ the combination $Q+F$ is always
even while for exotic hadrons with mass $O(n^2)$ the sum $Q+F$ is odd.

The Goldstone-Wiczek (GW) method is not an exact procedure. It becomes exact only in the limit where the soliton is 
so large that the variations of the fields of which is made out, can be considered adiabatically.
To compute exactly the fermion induce number on a soliton background, the complete analysis of the Dirac operator and its spectrum should be made (see for example \cite{Jackiw:1975fn,Niemi:1984vz}).
One of the first signals of the breaking of the GW technique is a zero mode crossing in the Dirac spectrum.

The large-$n$ limit is a weak coupling limit for ${\cal L}_{\rm eff}$, but  {\it is not} the limit in which the Goldstone-Wilczek approximation can be considered exact. 
In the large-$n$ limit the soliton size is constant, i.e. of order $n^0$, and the coupling constant is decreasing like $1/n$. 
The validity or not of the GW method, is encoded in the relation between the quadratic and the quartic term of the effective Lagrangian; these are the quantities that determine the size of the soliton. We also said that a zero mode crossing can change the quantum numbers of the soliton, but do not alter the conclusion about the $\Z_2$ stability.

\subsection{Generic Number of Flavor}
\label{general}

The purpose of the present section is to complete the study for generic values of $n_f$, giving a brief review of the results of \cite{Auzzi:2008hu}. We shall skip a lot of technical, but important, details that can be found by the reader in the given reference.

First of all, we need to explain the crucial difference between $n_f=2$ and higher $n_f$,
and the reason why the generalization is not a trivial task.

The residual symmetry, which was U(1) in  the $n_f =2$ case,
is now replaced by $\mathrm{SO}(n_f)$ with $n_f =3,4,5$.
All particles from the physical spectrum must thus
be classified according to representations of $\mathrm{SO}(n_f)$.
One can argue that for  $n_f =3,4,5$
the  Goldstone--Wilczek mechanism
provides the Skyrmion with an anomalous fermion number (and we shall see that this is the case). 
But this is not enough to guarantee the stability.  
For $n_f$ odd, we face a problem. 
This is due to the
existence of the antisymmetric tensor $\varepsilon^{i_1,i_2, ..., i_{n_f}}$
in $\mathrm{SO}(n_f)$. 
Using this tensor, we can
assemble $n_f$ composite fermions $\psi$ in a combination invariant under
the flavor group $\mathrm{SO}(n_f)$, creating a baryonic final state. For $n_f$ odd, this
state would have the
same quantum numbers as the Skyrmion, and thus, we could conclude that the
Skyrmion,  being an object with mass $\propto n^2$ would decay into
$n$ composite fermions $\psi$ with mass $O(n^0)$ in the flavor singlet configuration.
The anomalous fermion number, alone, does not imply the $\Z_2$ stability we are looking for.

An important role shall be played by
the spin and statistics, and the topological WZNW term that determines it.

Let us briefly remember
the situation in conventional QCD with fundamental quarks, where everything is clearly understood.
Also here there is an important difference between $n_f=2$ and higher $n_f$.
To begin with, let us consider two flavors, $n_f=2$.
In this case, the low-energy
chiral Lagrangian
does not admit the WZNW term, since $\pi_4(\SU(2)=\Z_2$ is not trivial. It does support
 Skyrmions, however. After quantization, the  Skyrmion quantum numbers
 $(I,\,J)$
form the following tower of possible values: $(0,0)$,
$(1/2,1/2)$, $(1,1)$, $(3/2,3/2)$, etc.
Here $I$ and $J$ stand for isospin and spin, respectively.
In the absence of the WZNW term, Skyrmions can be treated as both
bosons and fermions. This is due to the fact that we may or may not add
 an extra sign  in the field configurations belonging to nontrivial maps of
$\pi_4(SU(2))$ (the mechanism first discovered in \cite{Finkelstein:1968hy}).

At $n_f \geq 3$,
the choice of the Skyrmion statistics (i.e., boson vs. fermion)
becomes unambiguous.  At $n_f \geq 3$, it is possible (in fact, necessary) to introduce the WZNW term
in the effective Lagrangian \cite{Witten:1983tw,Witten:1983tx}. This term,
which is absolutely  essential in the anomaly matching between the
ultraviolet (microscopic) and infrared (macroscopic) degrees of freedom, is responsible for
the spin/statistics assignment for Skyrmions.

A similar situation takes place
in adjoint QCD. With two flavors, the WZNW term does not exist
since $\pi_4({\rm SU}(2)/{\rm U}(1))=Z_2$.
Quantization \cite{Krusch:2005bn} gives us two possible towers of states: bosons and fermions.  In the effective low-energy theory, it is
impossible to decide in which of the two towers the Skyrmion lies.
Only considering higher $n_f$, and computing the impact of the WZNW on the spin/statistic,  we can answer   this question.
The answer will play a crucial role in
the explanation of the Skyrmion stability.

We shall find that the Skyrmions are stable since they are
the only particles with an odd relation between statistics
and fermion number. Namely, Skyrmions can be bosons with
fermion number one or fermions with the vanishing fermion number.
Therefore, Skyrmions cannot decay to any final state
consisting of ``normal" or ``perturbative" particles.

As is well known \cite{Witten:1983tx}, the SO$(n)$ gauge theory with $n_f$ Weyl fermions in the
vectorial representation has the same as in Eq.~(\ref{pater}) pattern of
the global symmetry breaking,
and is also described by a nonlinear sigma model with the target space $\mathcal{M}_{n_f}$.
Witten  proposed \cite{Witten:1983tx} that
the Skyrmions of this theory must be identified with objects
obtained by contracting the SO$(n)$ antisymmetric tensor
$\varepsilon_{\alpha_1 \ldots \alpha_{n}}$ with the color indices
of the vectorial quarks and/or the gluon field strength tensor (see  Eqs.~(\ref{62}) and below).
These objects are stable due to the quotient symmetry
 $\mathbb{Z}_2={\rm O}(n)/{\rm SO}(n)$,
which acts as a global symmetry group.
We shall use this analogy to have a consistent picture, and check our results.

The section is organized as follows.
In \ref{lowenergy}, we
 describe in detail the low-energy effective action,
parametrization of the manifold ${\cal M}_{n_f}$, and
introduce a coupling to baryons $\psi$.  
In \ref{SOgaugetheory}, we describe the relevance of our results for another theory
with the same global symmetry breaking pattern, SO$(n)$ QCD with $n_f$ Weyl fermions
in the vectorial representation.  In \ref{WZNW},
we discuss the determination of the WZNW term and calculation
of its coefficient through the anomaly matching and describes the effect of the WZNW term on the
spin/statistics and fermion number of the Skyrmion. In \ref{stability}, we discuss an anomalous term
responsible for
the shift of the  Skyrmion fermion number, which, in turn, guarantees
its stability.

\subsubsection{Low-energy effective action}
\label{lowenergy}

In order to generalize to the case $n_f>2$, one must express the coset (\ref{target})
 in a way that makes ``evident'' the action of the
$\mathrm{SU}(n_f)$ symmetry. 
In the case of ${\rm SU}(2)/{\rm U}(1)$,
it was easy since using the representation with the unit vector $\vec{n}$ makes
evident how it transforms under  ${\rm SU}(2)$ rotations. The fermion interaction also follows easily (\ref{lagrangiana}). However,
the $n_f =2$ case can be somehow misleading for generalization to higher $n_f$.

${\rm SU}(2)$ can be represented as the sphere $S^3$ in the
four-dimensional vector space generated by the identity and the Pauli
matrices $\sigma_i$. Intersecting this sphere with the hyperplane
generated by the Pauli matrices,
we get an $S^2$ that is in one-to-one correspondence with the coset space
SU(2)/U(1). 
Moreover, this intersection tells us exactly how the
${\rm SU}(2)$ symmetry acts on the coset; it is the space $\{ \vec{n}\}$ of
the unit vectors. 
Another possible way though, is to intersect the space with
the hyperplane of the symmetric matrices generated by
$1,\, \sigma_1,\, \sigma_3$. 
This is again a sphere $S^2$ and is again in
one-to-one correspondence with the coset manifold. There is no
contradiction with the symmetry properties since for ${\rm SU}(2)$ the
adjoint representation is equivalent to the two-index symmetric and
traceless representation.
This is a consequence of equivalence between the
fundamental and the antifundamental representations in SU(2).

To generalize this construction to
higher $n_f$, we have to use the symmetric matrices. The space we get is
in one-to-one correspondence with the coset $\mathcal{M}_{n_f}$ and is
an explicit realization of its symmetric properties under the action
of the $\mathrm{SU}(n_f)$ group. We have thus a two-index symmetric matrix
that can be saturated by the fermion bilinear
$\psi^{a\alpha}\psi^{b\beta}\epsilon_{\alpha\beta}$.

The proper mathematical way to describe this is by using the Cartan embedding. 
The general element of the quotient $\mathcal{M}_{n_f}=
\mathrm{SU}(n_f)/\mathrm{SO}(n_f)$ can be written
in a compact form as $U\cdot\mathrm{SO}(n_f)$, where $U$ is an $\mathrm{SU}(n_f)$ matrix
(different $U$ in $\mathrm{SU}(n_f)$ corresponds to the same
$\mathcal{M}_{n_f}$ element, modulo a right product with an arbitrary
$\mathrm{SO}(n_f)$ element). The map
\beq
U\cdot\mathrm{SO}(n_f) \rightarrow W=U\cdot U^t \,,
\label{mapm}
\eeq
where the superscript $t$ denotes transposition,
is well-defined on the quotient because for the $\mathrm{SO}(n_f)$ matrices
the inverse is equal to the transposed matrix.
Equation (\ref{mapm}) presents  a one-to-one map between $\mathcal{M}_{n_f}$
and the submanifold of
the matrices of $\mathrm{SU}(n_f)$, which are both unitary and symmetric.

The Lagrangian of the Skyrme model with the target space $\mathcal{M}_{n_f}$
can be computed by evaluating the Lagrangian of the $\mathrm{SU}(n_f)$ Skyrme model
on the symmetric unitary matrix $W$,
\bea
\mathcal{L} &=&  \frac{F_\pi^2}{4} \mathcal{L}_2+ \frac{1}{e^2} \mathcal{L}_4 \nonumber\\[2mm]
&\equiv & \frac{F_\pi^2}{4} {\rm Tr} \left(\partial_\mu W \partial^\mu W^\dagger\right)+
\frac{1}{e^2} {\rm Tr} \left[ (\partial_\mu W) W^\dagger,(\partial_\nu W) W^\dagger \right]^2
\,.
\label{lagretta}
\eea

\subsubsection{Gauged formulation}

In the $n_f=2$  case, there are two ways to parametrize the target space $S^2$.
One can use a vector $\vec{n}$ subject to the constraint $|\vec{n}|  =1$. This is the so-called O(3) formulation.
Another approach is the  $z_i$ formulation where it is possible to express the Hopf charge (the charge of  $\pi_3(S^2)=Z$) as a local function of the gauge field $A$. An equivalent local expression in terms of the $\vec{n}$ field is impossible.

Generalization to higher $n_f$ \textit{ is not} achieved by extending the doublet to a complex
$n_f$-plet. For $n_f=2$, this strategy works because ${\rm SU}(2)$ is equivalent to the
 sphere $S^3$. In order to generalize to higher $n_f$, we need to start with an $\mathrm{SU}(n_f)$
 sigma model and then gauge an $\mathrm{SO}(n_f)$ subgroup.
 Let us consider the exact sequence
\[ \ldots  \rightarrow \pi_3\left({\rm SO}(k)\right) \rightarrow \pi_3\left({\rm SU}(k)\right)
\rightarrow
 \pi_3\left({\rm SU}(k)/{\rm SO}(k)\right)\rightarrow \pi_2\left({\rm SO}(k)\right)
\rightarrow \ldots \]
For every $k$, we have  $\pi_2\left({\rm SO}(k)\right)=0$. Therefore,
every non-zero element of $\pi_3\left({\rm SU}(k)/{\rm SO}(k)\right)$
can be lifted to a non-zero element of $ \pi_3\left({\rm SU}(k)\right)$
(for $n_f>2$ this lifting is not unique, as we will discuss below).
Then we can calculate the $S^3$ winding number
of the lifted 3-cycle,
using the $\mathrm{SU}(n_f)$ result,
\beq
s=-\frac{i}{24 \pi^2}  \int_{S^3} {\rm Tr} \, (U^\dagger dU)^3 \rule{0mm}{8mm}\,.
\eeq

It is possible to present the topological winding number as an
${\rm SU}(n_f)$ Chern--Simons current.
Let us introduce
\beq \mathcal{A}_\mu=i U^\dagger \partial_\mu U\,.
 \eeq
Then
\beq s=\frac{1}{8 \pi^2} \int d^3 x K^0, \,\,\,\,
K^\mu=\epsilon^{\mu \nu \rho \sigma} \,
\mathrm{Tr} \left(\mathcal{A}_\nu \partial_\rho \mathcal{A}_\sigma-
\frac{2}{3}\,i \, \mathcal{A}_\nu \mathcal{A}_\rho \mathcal{A}_\sigma\right).
\label{chs}\eeq
As previously discussed, $s$ is defined modulo $4$ for $n_f=3$
and modulo $2$ for $n_f>3$, due to arbitrariness in the choice of $U$.

\subsubsection{ The Fermion interaction}

Let us consider an $\mathrm{SU}(n_f)$ representative $U$
of a quotient class in $\mathcal{M}_{n_f}$.
The $\mathrm{SU}(n_f)$ symmetry group acts on $U$ as
$U \rightarrow R\cdot U$.
The action on the Cartan embedding image ($W=U\cdot U^t$) is
\begin{equation}
W \rightarrow R\cdot W\cdot R^t\,.
\end{equation}
Due to this property, we can write down the fermion coupling
for arbitrary $n_f$ as
\begin{equation}
-\frac{g}{2}\left\{ W^{fg} \psi_{\alpha f} \psi^\alpha_{g}+ \mathrm{h.c.}
\right\}.
\end{equation}

To the lowest order, the effective Lagrangian that includes both
pions and the fermions $\psi_{\alpha a}$ is
\beq
\mathcal{L}=\frac{F_\pi^2}{4} {\rm Tr} \, (\partial_\mu W \partial^\mu W^\dagger)+
\bar{\psi}_{f \dot{\alpha}} i \partial^{\dot{\alpha} \alpha} \psi_{f  \alpha}
-\frac{g}{2}\left\{ W^{fg} \psi_{\alpha f} \psi^\alpha_{g}+ \mathrm{h.c.} \right\}.
\label{leffe}
\eeq

If we expand around the vacuum where $W$ is given by the identity matrix,
the fermionic part of the Lagrangian is given by
\beq  \mathcal{L}_{\rm ferm}=
\bar{\psi}_{f \dot{\alpha}} i \partial^{\dot{\alpha} \alpha} \psi_{f  \alpha}
-g \left\{ \psi^\alpha_f \psi_{\alpha f} + \mathrm{H.c.} \right\}.\eeq
Of course, there are
 interactions between these fermions and the Goldstone bosons.

\subsubsection{Skyrmions in \boldmath{$\mathrm{SO}(n)$} QCD}
\label{SOgaugetheory}

Now we consider another parental theory: SO$(n)$ gauge theory with
$n_f$ Weyl quarks in the vectorial representation.  Such a theory
can be viewed as a ``parental"  microscopic theory because it has
the chiral symmetry breaking
\begin{equation}
\mathrm{SU}(n_f)\times \mathbb{Z}_{4 n_f}\to \mathrm{SO}(n_f) \times
\mathbb{ Z} _{2} \,, \label{paterso}
\end{equation}
which, apart from the discrete factors, is the same as SU$(n)$
Yang--Mills with adjoint Weyl quarks, see Eq.~(\ref{pater}).

The low-energy effective Lagrangian is again a nonlinear sigma model
with the target space
 ${\cal M}_{n_f}$.
The ``baryon number" symmetry,
which rotates all charge-$1$ Weyl quarks  is also anomalous;
the anomaly-free part is $\mathbb{Z}_{4 n_f}$.
This discrete symmetry is then broken down to $\mathbb{Z}_2$ by the fermion condensate.

There are
some differences from adjoint QCD.
One of them is that the coupling constant $F_{\pi}$
scales as $n$ rather than $n^2$. This means, in turn,  that now the
Skyrmion soliton is an object whose mass  scales as $n$.
Moreover, the fermion $\psi$ (see Eq.~(\ref{mafe})) is absent
in the spectrum.

The Skyrmion in the SO$(n)$ theory had been already  matched with
the stable particle construction  in the microscopic theory.
This identification belongs to Witten \cite{Witten:1983tx}. He
argued that the  Skyrmion corresponds to the baryon constructed of
$n$ quarks,
\begin{equation}
\label{barionSO} \epsilon_{\alpha_1 \alpha_2 \dots \alpha_{n}}
q^{\alpha_1} q^{\alpha_2} \dots q^{\alpha_{n}}\,.
\label{62}
\end{equation}

As was discussed in Ref.~\cite{Wittenbaryonvertex},
the gauge theory actually has an $\mathrm{O}(n)$ symmetry;
the quotient $\mathbb{Z}_2=\mathrm{O}(n)/\mathrm{SO}(n)$
acts as a global symmetry group.
All particles built with the $\epsilon_{\alpha_1 \alpha_2 \dots \alpha_{n}}$
symbol are odd under this symmetry. This means
that the baryon (\ref{62}) is stable under decay into massless
Goldstone bosons while two baryons can freely annihilate.

From Eq.~(\ref{62}),
 we can infer information about other quantum numbers of the
Skyrmion. Its $\mathbb{Z}_2$ fermion number is given by $n$ modulo $2$, and
 its flavor representation is contained in the tensor product of $n$ vectorial
representations. This is consistent with the computation carried out in conventional
QCD (with fundamental quarks)
in Ref.~\cite{Balachandran:1982dw}.

By the same token, we can argue that there is a similar contribution
to the fermion number of the Skyrmion in adjoint QCD,
which is proportional to $n^2-1$.
As discussed in Sect.~\ref{stability}, the composite fermion $\psi$
(which is absent in the SO$(n)$ theory)
will give an extra contribution to the Skyrmion fermion number, shifting it
by one unit.

\subsubsection{WZNW term}
\label{WZNW}

We can write the WZNW term for the $\mathcal{M}_{n_f}$ sigma model ($n_f\geq 3$)
by virtue of evaluating the ${\rm SU}(n_f)$ Wess--Zumino--Novikov--Witten term on the symmetric unitary matrices $W$
 introduced in Eq.~(\ref{mapm}). namely,
\bea
\Gamma
&=&
 -\frac{i}{240 \pi^2} \int_{B_5} d\Sigma^{\mu \nu \rho \sigma
\lambda} \mathrm{Tr} \left[ (W^\dagger \partial_\mu W)\cdot (W^\dagger
\partial_\nu W) \right.
\nonumber\\[3mm]
&\cdot&
\left. (W^\dagger \partial_\rho W) \cdot (W^\dagger \partial_\sigma W)\cdot
(W^\dagger \partial_\lambda W) \right].
\label{weszu}
\eea
In order to compute the WZNW term for the $\mathcal{M}_{n_f}$ sigma model,
we need  to take the result for $\mathrm{SU}(n_f)$
and restrict it to the submanifold of the unitary symmetric matrices.

There is a subtle difference regarding the possible coefficients allowed
for $\Gamma$ in the action.
In the Lagrangian of the $\mathrm{SU}(n_f)$ sigma model, relevant for
 QCD Skyrmions, the WZNW term must have just integer coefficient $k$,
\begin{equation}
\mathcal{L}= \mathcal{L}_2+ k \, \Gamma + \rm{Higher \, order \, terms}\,.
\end{equation}
 This is due to the fact that the integral of this term
on an arbitrary $S^5$ submanifold of $\mathrm{SU}(n_f)$ must be an integer multiple of $2 \pi$.
In the $\mathcal{M}_{n_f}$ sigma model relevant
for adjoint QCD,  we need to use the same normalization prescription.
The main difference is that if we integrate $\Gamma$ on
the minimal $S^5$, which we can build inside
the $\mathrm{SU}(n_f)$ subspace of the symmetric Hermitian matrices, the result will be
$4 \pi$ rather than $2 \pi$, as we get for the generator of $\pi_5\left(\mathrm{SU}(n_f)\right)$.
Therefore, if we restrict ourselves to this subspace
it is consistent to also consider half-integer values of $k$.

Let us gauge the U(1) subgroup generated by
\begin{equation}
Q=\pmatrix{ 0 & i & 0 \cr -i & 0 & 0 \cr 0 & 0 & 0 }.
\end{equation}
Let us take
\begin{equation}
T_{\kappa_1}=\pmatrix{ 1 & 0 & 0 \cr 0 & 1 & 0 \cr 0 & 0 & -2 }\,,
\end{equation}
which corresponds to the Goldstone boson $\pi_3$. We then find
\begin{equation}
\langle\partial_\mu J^\mu_{\kappa_1}\rangle =\frac{n^2-1}{16
\pi^2} \epsilon^{\kappa \nu \lambda \rho} F_{\kappa \nu} F_{\lambda
\rho}\,.
\end{equation}
At this point,
we can  match this value with the one found from the
low-energy theory.
We obtain in this way that the coefficient in
front of the WZNW term is
\begin{equation}
k=\frac{n^2-1}{2}.
\end{equation}
The crucial $1/2$ factor comes from the fact that
we  consider a theory with the Weyl fermions rather than
Dirac fermions as is the case in QCD.
Note that $k$ is half-integer for $n$ even
and integer for $n$ odd.

Using the arguments discussed previously,
it is straightforward to compute the coefficient $k$ of the WZNW term in the
low-energy effective action. The triangle diagram
is completely similar to that in the adjoint QCD case. The coefficient comes out
different
due to a different number of ultraviolet degrees of freedom.
The result  is
\begin{equation}
k=\frac{n}{2}.
\end{equation}
It immediately follows that  for $n$ odd the Skyrmion is a fermion
while for $n$ even it is a boson.
\begin{figure}[h!tb]
\epsfxsize=4.5cm \centerline{\epsfbox{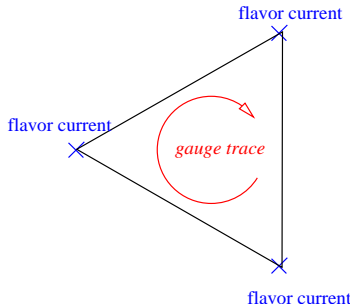}}
\caption{{\footnotesize The WZNW term is responsible for anomaly
matching between the ultraviolet (microscopic) theory and the low-energy effective
Lagrangian (macroscopic description). The anomalies
in question are given by  triangle graphs symbolically depicted in this figure, with flavor
currents in the vertices; they are blind with respect to the gauge indices. The only
information about the gauge structure comes from the trace in the
loop. }} \label{anomalyfig}
\end{figure}


Next, we have  to evaluate the WZNW term (\ref{weszu}) on the rotation.
The result is two times larger than that we get in QCD,
\beq \Gamma= 2 \pi\,.
\eeq

With our conventions for the coefficient
$k$, it can be an integer or half-integer, depending on the number
of colors $n$.
$k$ is half-integer for $n$ even
and an integer for $n$ odd.
It immediately follows that the Skyrmion is quantized as a fermion for
$n$ even and as a boson for $n$ odd.

\subsubsection{ Skyrmion stability due to anomaly}
\label{stability}

This section is central for the understanding of the
Skyrmion stability in the microscopic theory.

In order to generalize to higher $n_f$, one must consider
the triangle anomaly $${\rm U}(1)-\mathrm{SO}(n_f)-\mathrm{SO}(n_f)\,.$$
The U(1) corresponds to  the fermion number. For $\mathrm{SO}(n_f)$ we introduce an
auxiliary gauge field. The anomaly is
\begin{equation}
\partial_{\mu}J^{F0}_{\mu}=\frac{1}{16 \pi^2} \mathrm{Tr}(F^{\mu\nu}
\widetilde{F}_{\mu\nu}) = \frac{1}{8 \pi^2} \partial_\mu K^{\mu},
\end{equation}
where $F_{\mu\nu}=F^k_{\mu\nu} T^k$, with $T^k$ standing for the generators of
$\mathrm{SO}(n_f)$ (with ${\rm Tr} (T_jT_k)=\delta_{ij}$),
and $K_\mu$ is given in Eq.~(\ref{chs}).

The net effect of the baryon $\psi$ with mass $O(n^0)$ is to shift the
Skyrmion fermion
number by one unit, without changing its statistics.
For $n$ odd, the Skyrmion is a boson with an odd fermion number.
For $n$ even, it is a fermion with an even fermion number.
The relationship between the Skyrmion statistics and fermion number is abnormal.
In both cases, it is a $\mathbb{Z}_2$-stable object,
because in the ``perturbative" spectrum the normal relationship
between the fermion number and statistics takes place.

\section{Conclusion}

In orientifold QCD, we faced the problem of matching the Skyrmion with the right baryonic state. The minimal number of quarks, with which one
can build a particle with all quarks in the S-wave state, is $n(n \pm 1)/2$.
This $\sim n^2$ quark particle is the stable state described by the
Skyrmion. As for the $n$-quark particles, they are unstable with
respect to fusion of $n$ species into one $\sim n^2$ quark state, with a release of energy in the form of pion emission.

In Section \ref{stablebaryons}, we have made use of a certain
graphical representation of the baryons, in order to facilitate the
construction, and the study, of the gauge invariant wave functions. A graph is
in general formed by two components: the central elements  and the
lines that connect them. In the graphs previously used, the quarks were the central elements, and
the epsilon antisymmetric tensors were the lines connecting the quarks.

Now we want to introduce a ``dual'' graph, where the
central elements are the epsilon antisymmetric tensors, or baryon vertices, and the lines are the
quarks. All the properties previously derived can also be easily re-derived also with this dual formulation. We shall conjecture that these graphs have also a
physical realization in string theory.

\begin{figure}[h!tb]
\begin{center}
\leavevmode \epsfxsize 5 cm \epsffile{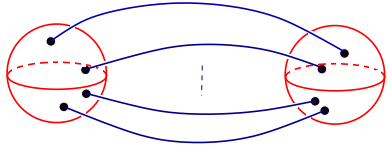}
\end{center}
\par \caption{\small Graph of the simplest baryon. Two baryon vertices connected by $n$ quarks.}
\label{basico}
\end{figure}

The graph for the simplest baryon Eq.~\ref{firstguess}   is
given in Figure \ref{basico}. The spheres
correspond to the baryon vertices, and the lines, $n$ of them,
correspond to the quarks. The epsilon tensors have $n$ indices, and
this implies that $n$ lines depart from every baryon
vertex. The quarks have two indices, and this corresponds to the fact
that a line in the graph connects two baryon vertices.

The other important baryons  are those  identified with the
Skyrmions of the low-energy effective Lagrangian. In the case of the
(S) representation, they consist of $n(n+1)/2$ quarks whose
indices are saturated by $n+1$ epsilon tensors. The number of quarks
is exactly equal to the dimension of the two-index symmetric
representation;  every
quark is living in a different state of the representation, and all the
states are occupied. The graph is composed by $n+1$ baryon
vertices, each of them connected once and only once with every other
baryon vertex by a quark line.  In Table \ref{table}, the right
column corresponds to the graphs for the baryon, respectively for
$n=2$ and $n=3$.
\begin{table}[h!tb]
\begin{center}
\begin{tabular}
[c]{|c|c|c|}\hline  & A
 &S \\ \hline \rule{0mm}{6mm} $n=2$ & \leavevmode
\epsfxsize 1.65 cm \epsffile{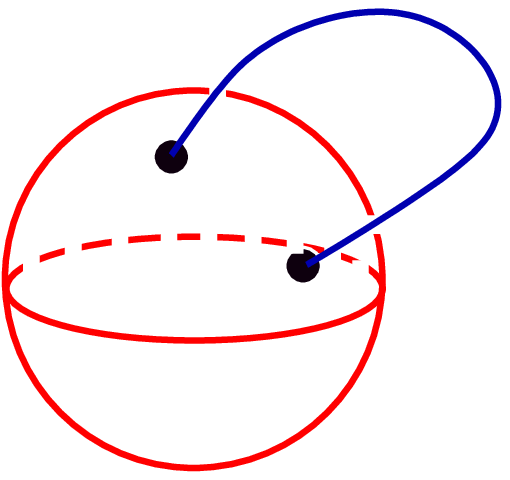}
 &\leavevmode \epsfxsize
4.7 cm \epsffile{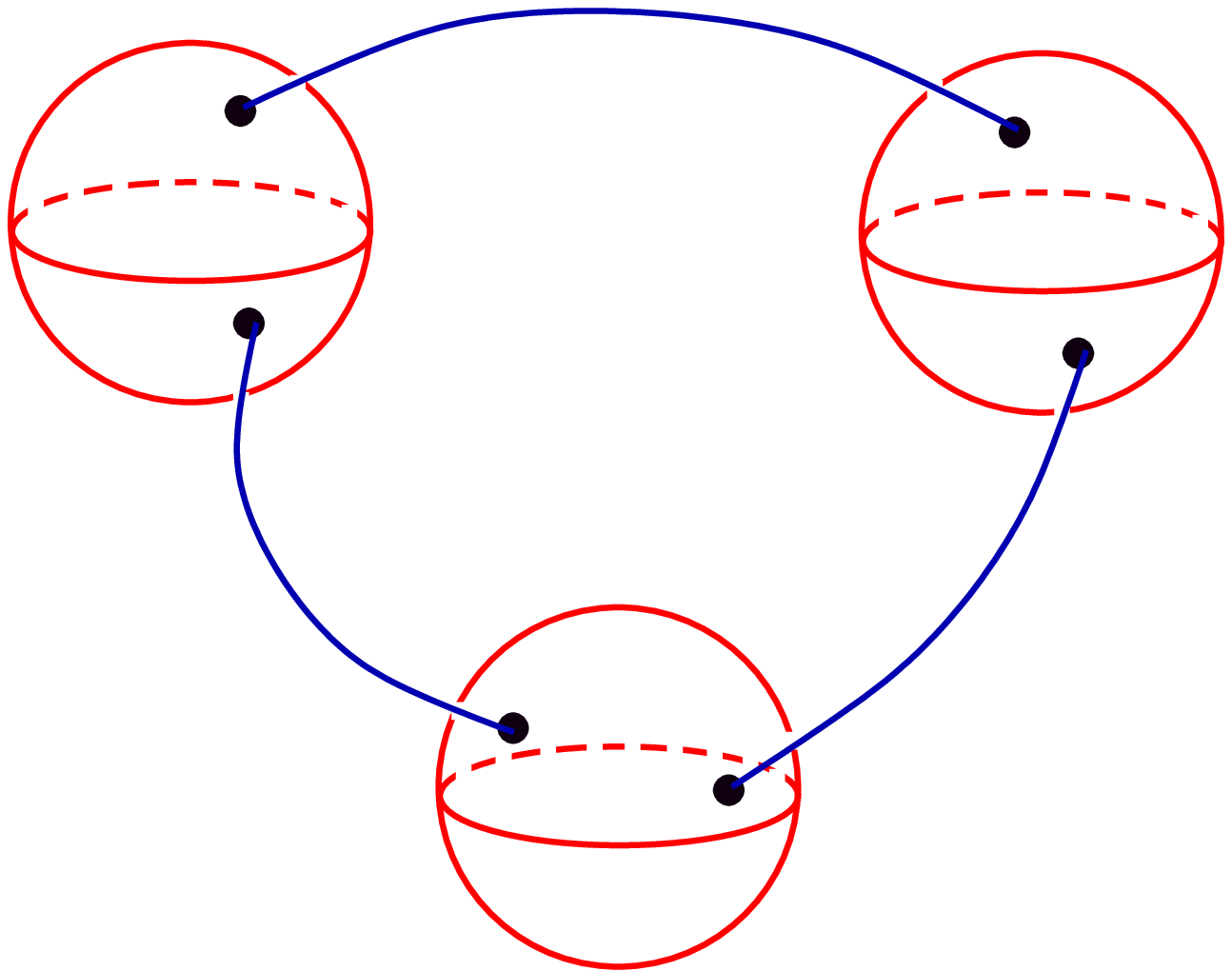} \\
\hline\rule{0mm}{6mm} $n=3$ & \leavevmode \epsfxsize 4.7 cm
\epsffile{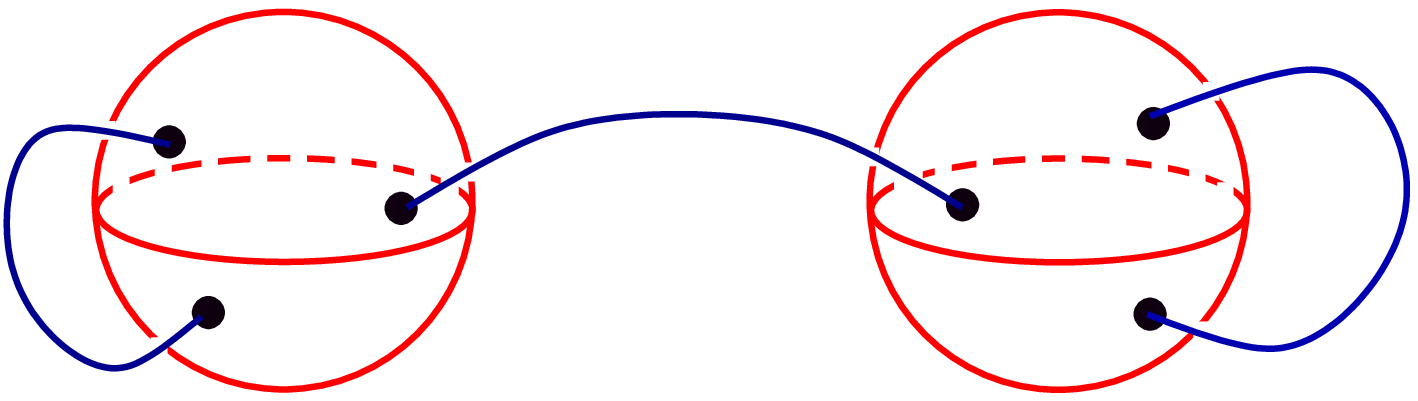}
 &\leavevmode \epsfxsize
4.9 cm \epsffile{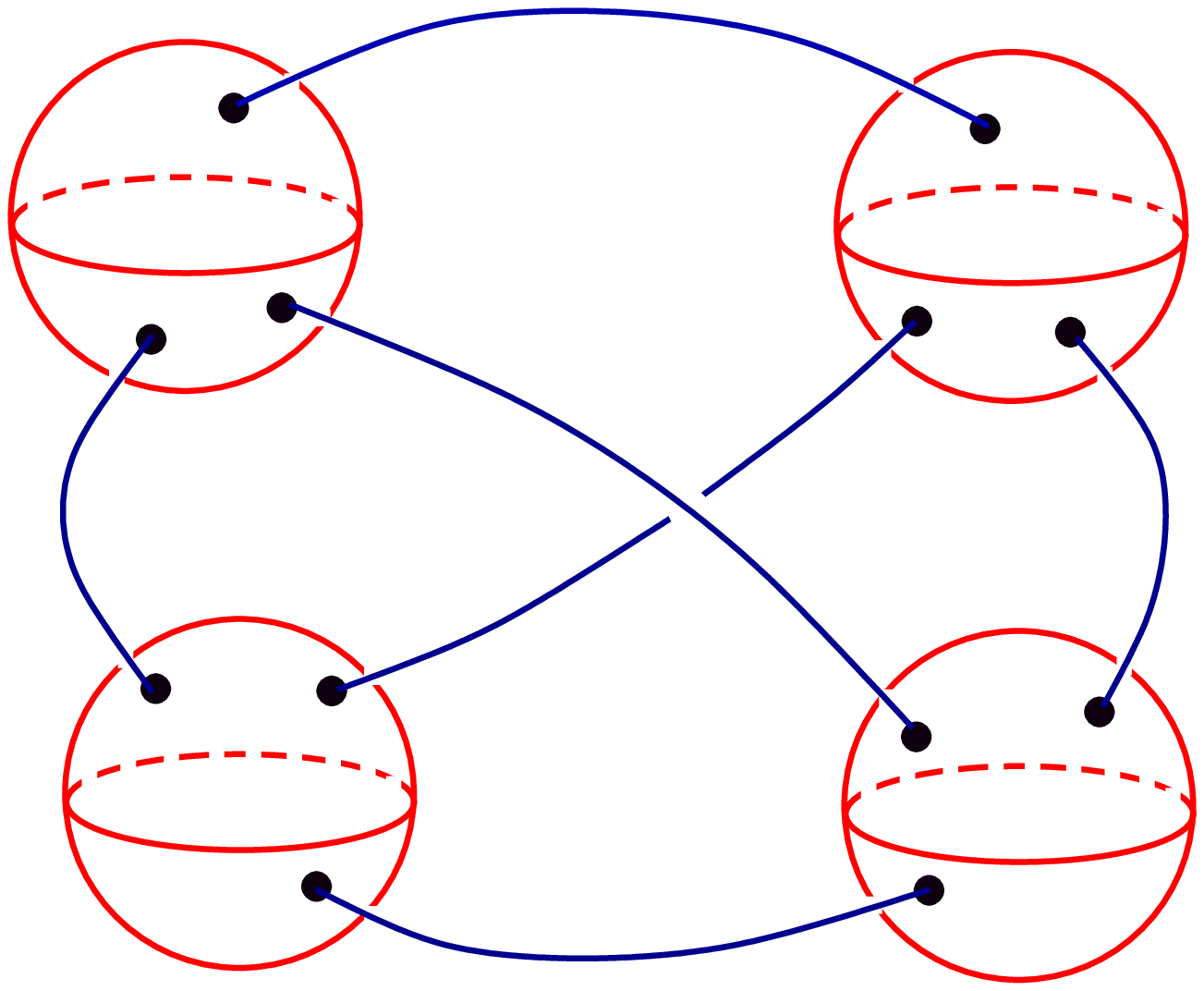} \\ \hline
\end{tabular}
\end{center}
\caption{\small Graphs of the Skyrmions. } \label{table}
\end{table}
The important property (Proposition \ref{proposition1}), is that there
is one and only one gauge wave function that is a gauge singlet, and
completely antisymmetric under the exchange of two quarks.  It can be easily understood from the use of these dual graphs. Every baryon vertex is connected to each one of the $n$ remaining baryon vertices by one and only one quark. Two quarks connecting the same two baryon vertices would spoil the required antisymmetric property.

In the case of the
(A) representation, the Skyrmion consists of $n(n-1)/2$ quarks
whose indices are saturated by $n-1$ epsilon tensors. The number of
quarks is exactly equal to the dimension of the two-index
antisymmetric representation.
The (A) representation is given in the left column of Table
\ref{table} for $n=2$ and $n=3$. The main difference with respect to
the symmetric representation is that now the two indices of one
quarks can be saturated in the same epsilon tensor. In the graph, the epsilon tensor is a line that starts and finishes at the same baryon vertex. In the
case of $n=2$, the quark is  already a singlet of the gauge group,
and the baryon is the quark itself. 
In the case of $n=3$, it
is given by Eq.~(\ref{treantisymmetricbaryon}.
The important property (Proposition \ref{proposition2}), is that there is one and only one gauge wave function that is a gauge singlet
and completely antisymmetric under the exchange of two quarks. This wave
function is composed by $n-1$ baryon vertices, connected together by $n(n-1)/2$ quarks
$Q^{[\alpha\beta]}$.  Every baryon vertex is connected to each one of the $n$ remaining baryon vertices by one and only one quark. In addition, every baryon vertex has a quark that starts and ends on itself.

\begin{figure}[h!tb]
\begin{center}
\leavevmode \epsfxsize 9 cm \epsffile{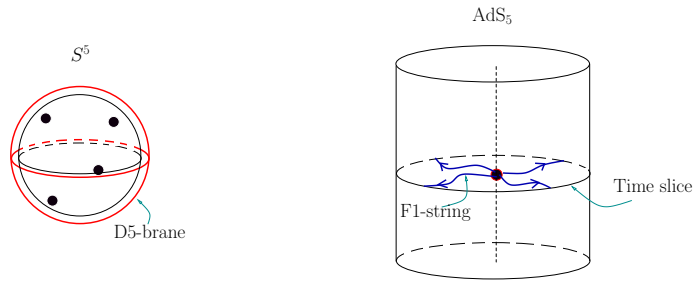}
\end{center}
\par \caption{\small Baryon vertex in the string dual of $\N=4$ SYM.}
\label{baryonvertex}
\end{figure}

Now moving to the string theory side, let us start explaining what is the baryon vertex in the
prototype of gauge-gravity duality: $\N=4$ $\SU(n)$ SYM. The string dual is
Type IIB on $\AdS_5 \times S^5$ with $n$ units of the RR flux passing
through the $S^5$ sphere. The baryon vertex (see Figure
\ref{baryonvertex}) consists of a D$5$-brane wrapped around the $S^5$
sphere \cite{Wittenbaryonvertex}.
The essential fact is a Chern-Simons term on the low-energy
effective action of the D$5$-brane $\int d^6x A \wedge G_5$ This is
an interaction between the bulk RR form and the gauge field living
on the brane. Integrating by part, it becomes $\int d^6x F \wedge
C_4$. This means that the RR form flux acts as a source for the brane
gauge field. In our case, the D$5$-brane is wrapped on a compact
manifold, and so the net charge for the gauge field must be zero. The
$n$ units of the RR flux must then be compensated by $n$ fundamental
strings ending on the brane. In the $\AdS_5$ side, the baryon vertex
lies in the $R=0$ line (the IR), and the $n$ strings that end from it
have the other extremity on the UV boundary, where eventually they
meet a probe quark source.

The string realization of orientifold QCD, and its gauge-gravity dual, is realized in the context of Type 0B string (see \cite{Angelantonj:1999qg,DiVecchiaReview} and reference therein\footnote{In \cite{Armoni:2008gg} there is a non-critical string realization.}). 
Here it is essential, in order to obtain the desired two-index representation for the quarks, to introduce a particular kind of orientifold plane. when a string crosses the orientifold, it changes the sign of its arrow and so it is possible to reproduce a quark with two indices both in the fundamental, or both in the antifundamental representation.  The gravitational side of the duality has no more orientifold plane; geometry keeps information about it. There are now fermionic, unoriented strings and bosonic, oriented ones.\footnote{I thank A.~Armoni for elucidations on this point.}

The idea, for the moment just a conjecture, is that the dual graphs presented before, may be the realizations of the baryons in the dual orientifold string theory.
If the duality works, clearly the gravity side must have a concrete realization of these baryonic states. 
The dual graphs, constructed out of baryon vertices and quarks, have a natural realization in string theory. Every baryon vertex should be a D$5$-brane wrapped in the internal space, and every quark should be a fermionic, unoriented, string.
The mechanism that selects the number of strings that can depart, or finish, on a baryon vertex, is still not clear. In $\N=4$, this was just the flux of the RR form. Here it must be something different, since fermionic strings are not oriented, and do not carry any flux. It is an interesting problem, open for future research.

\vskip 0.50cm
\begin{center}
*  *  *
\end{center}

\noindent 
In the second part of the paper, we considered adjoint QCD. 
If the number of flavors is greater than one, we have a continuous chiral symmetry breaking and a low-energy effective Lagrangian with target space $
\mathcal{M}_{n_f} = \mathrm{SU}(n_{f})/\mathrm{SO}(n_{f}) $. This Lagrangian supports topologically stable Skyrmions, much in the same way of ordinary QCD. 
The third homotopy groups of the coset space, summarized in Table \ref{tabone}, are nontrivial for all the values of $n_f$.

From the point of view of the low-energy effective Lagrangian, the Skyrmion is a topologically stable particle and belongs, at all rights, to the spectrum of the theory. In the large $n$ limit, this effective theory becomes more and more weakly coupled, and thus we are tempted to believe its prediction. But confronting ourselves with the microscopic description (adjoint QCD), we face many challenging questions.

First of all, at the contrary of ordinary QCD and orientifold QCD, we now do not have a conserved baryon number that could be associated with the topological current of the effective Lagrangian. We have a global $\U(1)$ symmetry, but it is broken by the anomaly and furthermore by the condensate $\lambda\lambda$. Not only do we  not have a microscopic object, like an operator made out of $F_{\mu\nu}$ and $\lambda$, to be matched with the Skyrmion, but we also do not have any apparent simple explanation for its stability.  We are thus faced with the following questions. {\it Is the Skyrmion an artifact of the effective Lagrangian or a real particle that is part of the spectrum of the theory? And if it is stable, what are the conserved charges that prevents it from decaying into lower excitations?}

A good way to start is to take a look at a different theory, $\SO(n)$ gauge theory with $n_f$ quarks in the vectorial representation, which shares the same $\chi$SB and so the same effective Lagrangian with adjoint QCD. More is already known about the Skyrmions in the $\SO$ theory. 
As suggested in \cite{Witten:1983tx}, the Skyrmion {\it is not } an artifact of the low-energy effective Lagrangian. It is part of the spectrum and should be identified with the baryon $\epsilon_{\alpha_1 \dots \alpha_n}q^{\alpha_1}\dots q^{\alpha_n}$.
Note that now we deal here with the $\mathbb{Z}_2$ stability: a composite state built
of two Skyrmions is not stable. This is in agreement with the fact that
$\pi_3(\mathcal{M}_{n_f})=\mathbb{Z}_2$  for $ n_f\geq 3$. 
The reason for the Skyrmion stability is due to $\mathbb{Z}_2=\mathrm{O}(n_c)/\mathrm{SO}(n_c)$ or, say in a more algebraic way, is necessary to have two epsilon tensor in order to write them a as a sum and product of deltas. This means that Skyrmions (with mass $\sim n$) can decay into mesons (mass $\sim n^0$), only if they annihilate in couples. A Skyrmion alone is absolutely stable.
Although we do not expect the same microscopic interpretation for adjoint QCD, this parental theory suggests to us that the Skyrmion is not an artifact but has a $\Z_2$ stability of some sort.

We started facing the problem with the lowest number of flavors $n_f=2$. In this case, the low-energy is a $\mathcal{M}_{2} = \mathrm{SU}(2)/\U(1)=S^2$ sigma model. The Skyrmion of this theory  is a knot type soliton and can be interpreted as a closed vortex, stabilized by a twist of the internal $\U(1)$ modulus. 
The topological current cannot be written in pure $S^2$ variables. We need to consider the full $\SU(2)$ three sphere with the $\U(1)$ gauged. The Skyrmion current corresponds to a Chern-Simons term for this $\U(1)$.

Now we remember that the global $\U(1)$ flavor symmetry, equivalent to the fermion number, although broken by the anomaly and $\lambda\lambda$, has a residual unbroken $\Z_2$. This will be crucial in the quest for stability.
For all ``ordinary" hadrons that can be produced from the vacuum by local
currents,  determination of $(-1)^F$ is
straightforward. If we classify the states according to this $(-1)^F$ fermion number and the $(-1)^Q$ parity, where $Q$ is the $\U(1)$ symmetry remnants of the flavor $\SU(2)$, all the perturbative particles are $(+1,+1)$ or $(-1,-1)$.

Another crucial aspect is that for adjoint QCD, unlike all the previously considered theories, there is a hadron carrying baryon number (in the $(-1)^F$ sense ) with mass $\sim n^0$ that {\it does not} grows with $n$. The simplest operator we can construct is made out of one fermion and one gauge tensor: $\psi_{\beta\,\,f} \propto \mathrm{Tr}\left( \lambda^{\alpha}_f\,F_{\alpha{\beta}%
}\right)$. We argued that this baryon, the lowest fermionic excitation that carries a baryon number, should included in the effective Lagrangian, in order to find the correct quantum numbers of the Skyrmion.

We argued that  the Skyrmions are exotic in the sense
that they have an induced fermion number coinciding with the Hopf
number, so that unlike all ``ordinary" hadrons they are
characterized by negative $(-1)^{Q+F}$. This is the underlying
reason explaining their stability.
The Hopf Skyrmion can have charges $Q$ and $F$, respectively, $0$ and $1$, if
there is no fermion zero mode crossing in the process of evolution from the
topologically trivial background to that of the Hopf Skyrmion, or $1$ and $0$
if there is a fermion zero mode crossing.
In both cases, the lightest exotic hadrons represented by the Hopf Skyrmion
are stable. They cannot decay in any number of pions and/or pions plus
``ordinary" baryons with mass $\O(n^0)$.  Note that this is a $\Z_{2}$
stability. Two Hopf Skyrmions can annihilate and decay into an array of $\pi$
's and $\psi$'s. For nonexotic hadron excitations which can be seen in a
constituent model and have mass $\O(n^0)$, the combination $Q+F$ is always
even, while for exotic hadrons with mass $\O(n^2)$ the sum $Q+F$ is odd.

Unlike the ``pion" part of the Lagrangian, unambiguously fixed by symmetries,
the ``hadron" part (\ref{lagrangiana})
does not seem to be unique. Indeed, the theory
under consideration has infinitely many fermionic interpolating operators.
For example, any operator of the type  Tr$(\lambda \dots\lambda F \dots)$
with an odd
number of $\lambda$'s represents a baryon.
One could pose the question of completeness. We would like to
claim that the Lagrangian
(\ref{lagrangiana}) is complete in the sense
that it captures all relevant dynamics to answer the question we pose (the quantum numbers of the Skyrmions),
there is no need to include in it additional baryon operators.
The point is that any other baryons with mass $O(n^0)$
(which are necessary unstable particles, resonances)
have a projection on the operator (\ref{mafe}).
The inclusion of additional baryon terms would have no impact on our result
since they would not change the anomaly.

We then  advanced  the results that had been obtained previously
to three or more flavors.

The generalization is not a trivial task.
The residual symmetry that was U(1) in  the $n_f =2$ case
is now replaced by $\mathrm{SO}(n_f)$ with $n_f =3,4,5$.
For $n_f$ odd we face a problem. This is due to the
existence of the antisymmetric tensor $\varepsilon^{i_1,i_2, ..., i_{n_f}}$
in $\mathrm{SO}(n_f)$. Using this tensor, we can
assemble $n_f$ composite fermions $\psi$ in a combination invariant under
the flavor group $\mathrm{SO}(n_f)$, creating a baryonic final state. For $n_f$ odd, this
state would have the
same quantum numbers as the Skyrmion, and thus, we could conclude that the
Skyrmion,  being an object with mass $\propto n^2_c$ would decay into
$n_f$ composite fermions $\psi$ with mass $O(n^0_c)$ in the flavor singlet configuration.
The anomalous fermion number, alone, does not imply the $\Z_2$ stability we are looking for.

An important role  is now played by
the spin and statistics and the topological WZNW term that determines it.
At $n_f \geq 3$,
the choice of the Skyrmion statistics (i.e., boson vs. fermion)
becomes unambiguous.  At $n_f \geq 3$, it is possible (in fact, necessary) to introduce the WZNW term
in the effective Lagrangian \cite{Witten:1983tw,Witten:1983tx}. This term,
which is absolutely  essential in the anomaly matching between the
ultraviolet (microscopic) and infrared (macroscopic) degrees of freedom, is responsible for
the spin/statistics assignment for Skyrmions.

We find that the Skyrmions are stable since they are
the only particles with an odd relationship between statistics
and fermion number. Namely, Skyrmions can be bosons with fermion number one or fermions with the vanishing fermion number.
Therefore, Skyrmions cannot decay to any final state
consisting of ``normal" or ``perturbative" particles: pions
and other similar mesons or baryons of the type (\ref{mafe}).

The underlying (microscopic) reason for the Skyrmion stability
is an odd relationship between the Skyrmion  statistics and its fermion number.
For $n_c$ odd, the Skyrmion is a boson with an odd fermion number.
For $n_c$ even, it is a fermion with an even fermion number.

Our analysis is valid at large $n_c$.
Something peculiar happens when we leave the
large-$n_c$ limit and go to small $n_c$.
${\rm SU}(n_c)$ adjoint QCD  for $n_c=2$
and the SO$(n_c)$ gauge theory with vector quarks
 for $n_c=3$ are in fact one and the same theory.
 The SO$(n_c)$ description
in this particular case is better.
In this case, the fermion $\psi$ coincides with
the Pfaffian, $\epsilon_{abc} q^a F^{b c}$;
therefore, it does not make sense to introduce it as
another independent degree of freedom.

An interesting  issue  that remain to be clarified is whether or not the flux tubes
supported by the chiral Lagrangian are related to
confining strings of the underlying
microscopic gauge theory.
Note in fact that $ \pi_2 (\mathrm{SU}(n_f)/\mathrm{SO}(n_f)) = \mathbb{Z}_2$ for $n_f\geq 3$.
This fact implies that the sigma model does indeed support flux tubes. These flux tubes are $\mathbb{Z}_2$-stable,
i.e., a pair of them can annihilate.
The microscopic theory analyzed by Witten \cite{Witten:1983tx,Wittenbaryonvertex,benson-saadi} was
O$(n_c)$ gauge theory, with quarks in the vector representation
of O$(n_c)$.  
His argument, based on the fact that
an external probe quark in the spinor representation of O$(n_c)$
cannot be screened by dynamical quarks in the vector representation, was to identify these strings with the ones that confines spinorial probe quark.
Now we know that one and the same pattern of the chiral symmetry
breaking  takes place in the O$(n_c)$ gauge theory with
quarks in the vector representation {\em and} SU$(n_c)$ gauge theory
with quarks in the adjoint representation. However, Witten's
argument is totally inapplicable in the latter case. Indeed, in this
microscopic theory a probe quark with any number of, say, upper
indices $Q^{i_1 \,...\, i_n}$ and no lower indices cannot be
screened by adjoint dynamical quarks. Strings of any $n$-ality, up
to $[n_c/2]$, are stable. (Here [...] stands for the integer part.)
This tells us that the $\mathbb{Z}_2$-strings supported by the
chiral low-energy theory are unrelated to the confinement strings of
the corresponding microscopic theories. What phenomenon do they
describe?

Our analysis of the adjoint Skyrmion is not as complete as the one of the orientifold one. 
In the latter case,  we have a clear understanding of the Skyrmion,  both from the effective low-energy Lagrangian, as a soliton, and from the microscopic theory, as a baryon operator. 
Our identification has many, independent arguments in favor of it.
In the adjoint QCD case, the Skyrmion analysis is restricted to the low-energy Lagrangian. 
We found a reason to believe that the stability of this object is a true feature of the fundamental theory, and not an artifact of the low-energy approximation. We still do not have a clear understanding of what this particle represent in the microscopic theory. 
It could be a non-local object, that is not possible to write as a local combination of operators acting on the vacuum.
This is certainly one issue that has to be clarified in the future.


\section*{Acknowledgements}

The first part of the project began when I was in Copenhagen, founded by the Marie
Curie Grant MEXT-CT-2004-013510. I want to thank F.~Sannino for useful conversations there.
The second part of the paper has been carried out in Minneapolis at FTPI. I am grateful to R.~Auzzi for useful discussions and the collaboration \cite{Auzzi:2008hu} on which the last part of this contribution is based. 
I want to thank especially M.~Shifman for many useful discussions and collaborations
\cite{Bolognesi:2007ut,Auzzi:2008hu}.
I want also to thank A.~Armoni for comments on the manuscript.
This review grew out of a seminar given in September 2007 at the Isaac Newton Institute in Cambridge.  I want to people there for the invitation and for useful discussions.
My work is now supported by DOE grant DE-FG02-94ER40823.

\end{document}